\documentclass[amsmath, amssymb, twocolumn, aps, prx, superscriptaddress]{revtex4-1}

\usepackage{mathtools}
\usepackage{mathrsfs, amsmath, wasysym}
\usepackage{bbold}
\usepackage{dsfont}
\usepackage{braket,mleftright}
\usepackage[mathscr]{eucal}
\usepackage{xfrac}
\usepackage{graphicx}
\usepackage{dcolumn}
\usepackage{bm}
\usepackage{color}
\usepackage{tabularx}
\usepackage{natbib}
\usepackage[colorlinks,linkcolor=blue,citecolor=blue,urlcolor=blue]{hyperref}
\usepackage[normalem]{ulem}
\usepackage[all]{hypcap}
\usepackage[dvipsnames,usenames,table]{xcolor}

\newcommand{\bg}{\begin{gather} }
\newcommand{\eg}{\end{gather}}

\newcommand{\be}{\begin{equation} }
\newcommand{\ee}{\end{equation}}
\newcommand{\beq}{\begin{eqnarray}}
\newcommand{\eeq}{\end{eqnarray}}

\newcommand{\bK}{{\bf K}}
\newcommand{\bq}{{\bf q}}
\newcommand{\bx}{{\bf x}}
\newcommand{\by}{{\bf y}}
\newcommand{\bz}{{\bf z}}

\newcommand{\ba}{{\bf a}}

\newcommand{\bp}{{\bf p}}
\newcommand{\br}{{\bf r}}

\newcommand{\bS}{{\bf S}}

\newcommand{\bM}{{\bf M}}
\newcommand{\bQ}{{\bf Q}}

\newcommand{\brho}{\boldsymbol{\rho} }
\newcommand{\bdelta}{\boldsymbol{\delta} }

\newcommand{\bfell}{\boldsymbol{\ell} }

\def\black{\color{black}}

\begin{document}

\title{Excitonic instability towards a Potts-nematic quantum paramagnet}

\author{Jeremy Strockoz}
\affiliation{Department of Physics, Drexel University, Philadelphia, PA 19104, USA}

\author{Daniil S. Antonenko}
\thanks{J. Strockoz and D. S. Antonenko contributed equally to this work.}
\affiliation{Department of Physics, Drexel University, Philadelphia, PA 19104, USA}

\author{Dmitri LaBelle}
\affiliation{Department of Physics, Drexel University, Philadelphia, PA 19104, USA}

\author{J{\"o}rn W. F. Venderbos}
\affiliation{Department of Physics, Drexel University, Philadelphia, PA 19104, USA}
\affiliation{Department of Materials Science \& Engineering, Drexel University, Philadelphia, Pennsylvania 19104, USA}

\date{\today}

\begin{abstract}
Magnetic frustration can lead to peculiar magnetic orderings that break a discrete symmetry of the lattice in addition to the fundamental magnetic symmetries (i.e., spin rotation invariance and time-reversal symmetry). In this work, we focus on frustrated quantum magnets and study the nature of the quantum phase transition between a paramagnet and a magnetically ordered state with broken threefold ($\mathds{Z}_3$) crystal rotation symmetry. We predict the transition to happen in two stages, giving rise to an intermediate nematic phase in which rotation symmetry is broken but the system remains magnetically disordered. The nematic transition is described by the three-state Potts model. This prediction is based on an analysis of bound states formed from two-magnon excitations in the paramagnet, which become gapless while single-magnon excitations remain gapped. By considering three different lattice models, we demonstrate a generic instability towards two-magnon bound state formation in the Potts-nematic nematic channel. We present both numerical results and a general analytical perturbative formula for the bound state binding energy similar to BCS theory. We further discuss a number of different materials that realize key features of the model considered, and thus provide promising venues for possible experimental observation. 
\end{abstract}

\maketitle

\section{Introduction \label{sec:introduction}}

The study of ordering phenomena is one of the central pillars of condensed matter physics. Landau's theory of phase transitions provides a general framework for understanding the structure of ordered phases and the nature of the transitions between different phases. The order parameter is a fundamental concept in the study of phase transitions and characterizes the ordered phase in terms of its symmetry. In the simplest and most common cases, the ordering transition is associated with the breaking of a fundamental symmetry, such as translation symmetry in the case of crystalline order; charge conservation in the case of superconductivity; and time-reversal combined with spin-rotation symmetry in the case of magnetic order. Ordering transitions with a much richer structure arise when, in addition to the fundamental symmetry, one or more secondary symmetries are broken in the ordered state, such as spatial rotation or inversion symmetry. This occurs, for instance, in unconventional non-$s$-wave superconductors and superfluids, where the internal structure of the Cooper pairs (i.e., their spin and orbital angular momentum) can lead to spontaneous spatial anisotropy associated with broken rotation symmetry~\cite{Sigrist:1991p239,Leggett:1975p331}. Similar phenomena can occur in magnets when the exchange interactions are (strongly) frustrated, which prohibits the formation of simple ferromagnetic or antiferromagnetic N\'eel order. Whereas the ferromagnet and antiferromagnet do not break the symmetries of the crystal lattice, the more complicated magnetic configurations that arise as a result of frustration generally do break one or more spatial symmetries. 

\begin{figure}
\includegraphics[width=0.9\linewidth]{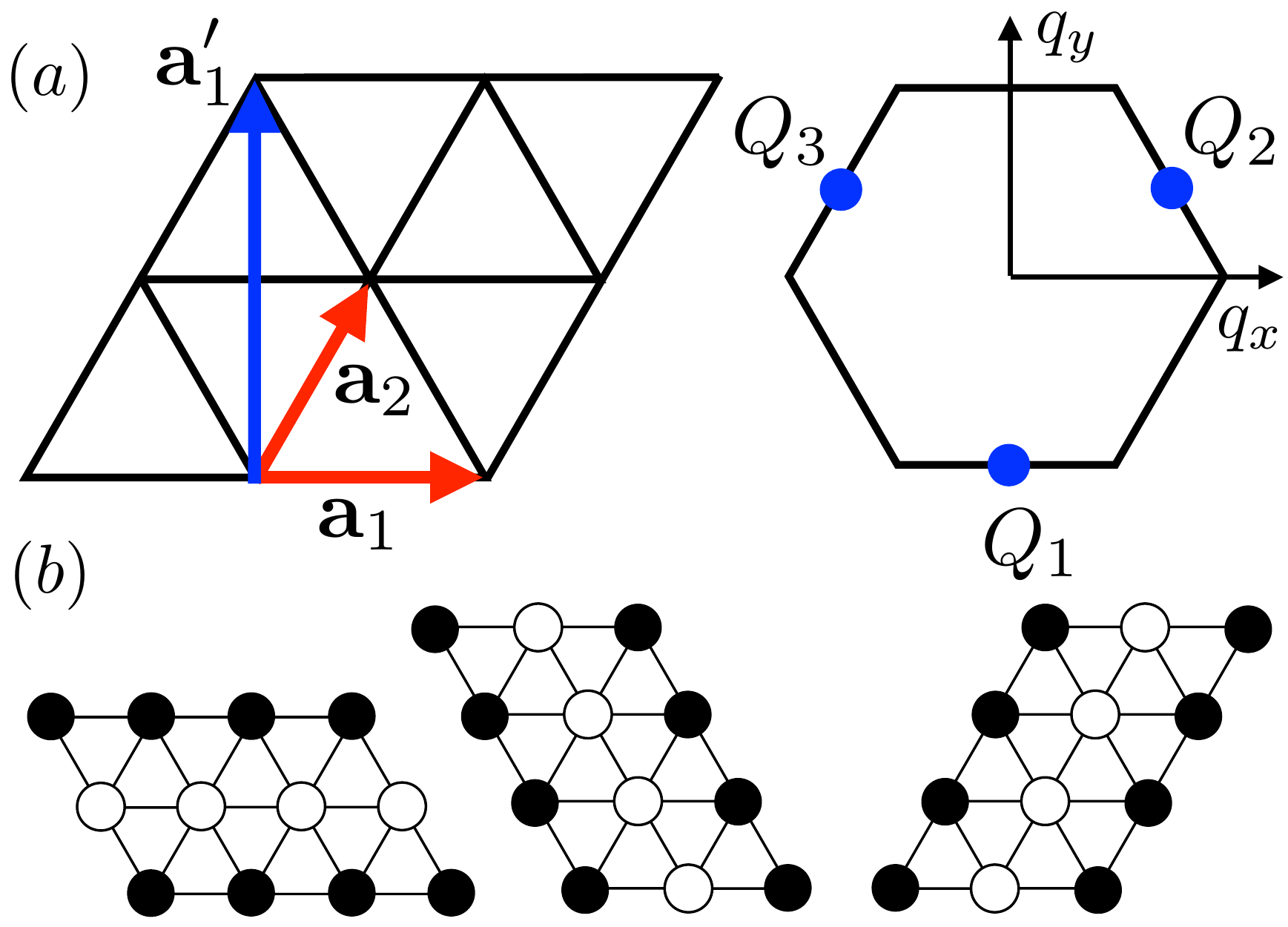}
\caption{{\bf Triangular lattice model.} (a) Triangular lattice Heisenberg model with exchange couplings $J_1$ between nearest neighbors (exemplified by red arrows) and $J_2$ between next-nearest neighbor (blue arrow). When both $J_{1,2}>0$ and $1/8 < J_2/J_1 <1 $, the minima of the Fourier-transformed exchange coupling matrix $J_\bq$ are located at the three $M$ points in the Brillouine zone with the wave vectors $Q_{1,2,3}$, indicated by blue dots on the right. The generating lattice vectors $\ba_{1,2}$ are indicated in red. (b) Three possible realizations of triangular lattice single-$Q$ ordering corresponding to the three $M$-point wave vectors. Black (white) solid dots represent classical moments pointing out of (into) the page. }
	\label{fig:triangular}
\end{figure}

In this paper, we study a particular class of such frustrated magnets: magnets which spontaneously break a discrete $\mathds{Z}_3$ (i.e., threefold) crystal rotation symmetry in the ordered state. Such phases naturally arise in simple Heisenberg spin models with frustrated exchange interactions, for instance in two-dimensional (or layered quasi-two-dimensional) magnets with triangular ~\cite{Jolicoeur:1990p4800,Korshunov:1993p6165,Lecheminant:1995p6647} or honeycomb lattice~\cite{Fouet:2001p241} geometry. The phase diagrams of triangular and honeycomb lattice spin models with further neighbor exchange couplings are known to include collinear single-$Q$ magnetic states, characterized by staggered ferromagnetic stripes (see Fig.~\ref{fig:triangular}), which are stabilized by the ``order by disorder'' mechanism \cite{Villain:1980p1263,Henley:1987p3962,Henley:1989p2056,Jolicoeur:1990p4800,Chubukov:1992p11137}. Since a single ordering wave vector is spontaneously selected from a set of three wave vectors related by crystal rotation symmetry, the latter symmetry is spontaneously broken in these single-$Q$ states. This may be compared to Heisenberg antiferromagnets with tetragonal symmetry, such as frustrated square lattice magnets, which can support single-$Q$ stripe order described by either one of two wave vectors $(\pi,0)$ and $(0,\pi)$ \cite{Henley:1989p2056,Chandra:1990p88}. Such single-$Q$ phases break a $\mathds{Z}_2$ Ising symmetry corresponding to the two possible orientations of the stripes, which in two dimensions implies a finite temperature Ising transition to an Ising nematic phase with no magnetic order \cite{Chandra:1990p88}. In three dimensions the Ising and magnetic transitions may still occur at different temperatures, in which case a so-called vestigial Ising nematic phase arises~\cite{Fernandes:2012p024534}. The emergence of Ising nematicity is well-known example of the rich ordering phenomenology that can occur in complex magnets with multiple broken symmetries~\cite{Fernandes:2019p133}.

In contrast to the Ising case, the breaking of a discrete $\mathds{Z}_3$ symmetry is governed by the three-state Potts model~\cite{Wu:1982p235} and thus provides a manifestly different---and much less studied---window into the nature of magnetic ordering in unconventional complex magnets~\cite{Tchernyshyov:2002p064403,Mulder:2010p214419,OrthArxiv}. The purpose of this paper is to explore the nature of the phase transition in magnets that break an additional $\mathds{Z}_3$ symmetry, such as a threefold crystal rotation, and thus belong to the three-state Potts universality class. As far as the thermal phase transition is concerned, general arguments suggest that, similar to the Ising case, a finite temperate transition to a Potts-nematic phase is possible~\cite{Mulder:2010p214419,OrthArxiv}. 

Rather than the thermal phase transition, our goal here is to study the nature of the zero temperature quantum phase transition between a quantum paramagnet and a magnetically ordered phase which breaks a discrete $\mathds{Z}_3$ crystal rotation symmetry. The question we specifically seek to answer is whether an intermediate Potts-nematic phase, in which the system remains paramagnetic but breaks the $\mathds{Z}_3$ rotation symmetry, can exist. 

To address this question, we consider a minimal XXZ model of $S=1$ quantum spins with single-ion anisotropy \cite{Wang:2017p184409}. In this minimal model, defined in Eq. \eqref{eq:H_XXZ}, the single-ion anisotropy $D$ controls the transition between the quantum paramagnet realized for sufficiently large $D$ and the magnetically ordered state. The latter is controlled by the frustrated exchange interactions, which are chosen such that, at the classical level, the energy is minimized by a set of degenerate ordering wave vectors related by threefold rotation symmetry. Following the approach of Ref.~\onlinecite{Wang:2017p184409}, we start from the quantum paramagnet and study its instability towards the formation of two-magnon bound states as $D$ is lowered, which may occur due to an attractive interaction between single-magnon excitations. If the energy gap of two-magnon bound states closes at a value of $D$ larger than the value at which the single-magnon gap closes, this indicates an instability towards bound state formation. Furthermore, if the internal structure of the bound state solution, defined by the relative angular momentum of the magnons, breaks the $\mathds{Z}_3$ rotation symmetry, this suggests that proliferation of two-magnon excitations gives rise to an intermediate Potts-nematic quantum paramagnetic phase. Such analysis is similar in spirit to the classic Cooper problem in the context of superconductivity~\cite{Cooper:1956p1189}, which addresses the instability of the Fermi sea to the formation of two-particle bound states (Cooper pairs). 

We apply this analysis to three different models of frustrated quantum magnets: the triangular lattice, the honeycomb lattice, and the three-dimensional face-centered cubic (FCC) lattice. In all three cases we focus on a part of the phase diagram where rotation symmetry broken single-$Q$ magnetic order is favored. We obtain a matrix Schr\"odinger equation for the two-magnon bound states, show that it decouples in channels of distinct symmetry quantum numbers, and solve it numerically. We demonstrate that when the exchange interactions favor magnetic order broken rotation symmetry, generically two-magnon bound states have largest (positive) binding energy in the nematic channel, suggesting a generic instability towards a Potts-nematic phase in this class of quantum magnets. The numerical analysis is supplemented by a perturbative analytical approach, is independent of the underlying lattice structure and in two dimensions yields an expression for the width of the intermediate Potts-nematic phase similar to BCS theory. In particular, as expressed in Eq.~\eqref{binding_energy}, the binding energy, which is proportional to the width of the nematic phase, depends on the single-magnon density of states and the effective coupling constant in the nematic channel. 

To make a connection with possible candidate materials, we identified a number of compounds that can host the proposed Potts-nematic vestigial phase under certain conditions. We point to two specific material classes, which we propose as a prospective platform to test our predictions. One of the two groups is the family of transition metal trichalcogenides ($\text{M}P\text{X}_3$), which have attracted much attention recently \cite{Wildes:1998p6417, Wildes:2012p416004, Wildes:2015p224408, Lancon:2016p214407, Kim:2020p184429, Calder:2021p024414, Ni:2022p3283}. The other is iron-intercalated transition metal dichalcogenide $\text{Fe}_x \text{Nb} \text{S}_2$, which was recently reported to host a $\mathds{Z}_3$-breaking magnetism, revealed in optical measurements \cite{Little:2020p1062,Wu:2022p021003}.

The paper is organized as follows. In the next Section~\ref{sec:Z3_breaking} we define our model and give general introduction to the physics of $\mathds{Z}_3$-symmetry breaking magnets.  We proceed with the Section~\ref{sec:theory}, in which we develop the main formalism used in our work and derive its perturbative limit.  We then apply this machinery to the specific cases of triangular, honeycomb, and face-centered cubic lattices and present the results in Section~\ref{sec:results}. Finally, we propose concrete experimental implementations in Section~\ref{sec:expts} and conclude with a discussion (Section~\ref{sec:discussion}). The details of the derivation for a non-Bravais lattice are relegated to Appendix~\ref{app:gen}.  Throughout the paper we consider zero temperature and work in units of the lattice constant (i.e., $a = 1$).

\section{Rotation symmetry breaking in frustrated antiferromagnets \label{sec:Z3_breaking}}

This work is focused on frustrated Heisenberg antiferromagnets described by the general Hamiltonian 
\be
H = \frac12 \sum_{ij} J_{ij} \bS_i \cdot \bS_j, \label{eq:H_HB}
\ee
where $J_{ij} = J_{ji}$ are the exchange coupling constants. When the spins $\{ \bS_i \}$ are considered as classical vectors, the exchange couplings describe interactions between pairs of spins which either favor alignment ($J_{ij}<0$) or anti-alignment ($J_{ij}>0$). Frustration arises when the interactions between different pairs of spins compete and the energetic requirement of perfect alignment or anti-alignment cannot be simultaneously satisfied for all pairs of spins. Frustration may originate from the geometry of the lattice, as is the case for the nearest neighbor triangular lattice antiferromagnet, or from the presence of multiple competing exchange constants, such as in $J_1$--$J_2$--$J_3$ models.

A standard approach to determining the classical spin configuration with the lowest energy is the Luttinger-Tisza method~\cite{Luttinger:1946p954,Luttinger:1951p1015}. It relies on expanding the spin variables in Fourier modes as $\bS_{\br \alpha} = \sum_{\bq} e^{i \bq\cdot (\br+\bfell_\alpha)} \bS_{\bq \alpha}$, where $\bfell_\alpha$ denotes the position of sublattice site $\alpha$ with respect to the lattice vector $\br$, and writing the spin Hamiltonian in terms of the Fourier modes as $H =  (N/2) \sum_{\bq} ( J_{\bq} )_{\alpha\beta}  \bS^*_{\bq \alpha} \cdot \bS_{\bq \beta}$. Here $ J_{\bq}$ is the Fourier transform of the exchange couplings, which in the case of non-Bravais lattices is matrix in sublattice space. Minimization of the classical energy then requires computing the eigenvalues of $J_{\bq}$ for all $\bq$ and finding the minimum of lowest eigenvalue branch with respect to $\bq$. The minimum occurs at some specific wave vector $\bQ$ or at a set of symmetry related (i.e., ``degenerate'') wave vectors $\{\bQ_i\}$. The latter case is a consequence and signature of frustration. The eigenmodes corresponding to the minima of $J_\bq$ are generally referred to as the optimal Luttinger-Tisza modes and can be used to construct spin states that minimize the magnetic energy. This construction must be subject to the fixed-length constraint, i.e., $|\bS_i|^2=S^2$, however, and this implies that, in the case of non-Bravais lattices, it may not be possible to construct ground states from optimal Luttinger-Tisza modes only. In this case, the Luttinger-Tisza method only provides a lower bound the classical energy. Note that in the case of primitive Bravais lattices, however, it is always possible to construct ground states from the optimal Luttinger-Tisza modes.

In this work we focus on a particular class of frustrated magnetic systems for which the minima of the eigenvalues of $J_\bq$ occur at a set of three symmetry-related wave vectors $\{\bQ_i\}$ (with $i=1,2,3$) in some part of the phase diagram. The degenerate wave vectors are related by crystal rotation symmetry and satisfy the two additive relations
\be
2 \bQ_i \simeq 0, \qquad \bQ_1+\bQ_2+\bQ_3 \simeq 0, \label{eq:Q-relation}
\ee
Note that the first of these relations implies that the Fourier modes are real. Prominent examples of Heisenberg magnets which exhibit such phases in the phase diagram are the triangular $J_1$--$J_2$ model~\cite{Jolicoeur:1990p4800,Chubukov:1992p11137}, the honeycomb lattice $J_1$--$J_2$--$J_3$ model~\cite{Fouet:2001p241}, and the three-dimensional face-centered cubic (FCC) lattice $J_1$--$J_2$ model~\cite{Oguchi:1985p4494,Balla:2020p043278,Henley:1987p3962}. These three models will be the subject of the remainder of this paper and will be discussed in more detail below. In the case of the triangular and honeycomb lattices, the wave vectors $\bQ_i$ correspond to the high symmetry $M$ points of the Brillouin zone shown in Fig.~\ref{fig:triangular}(a), which are clearly related by threefold rotation. The triangular lattice provides the simplest example of the magnetism of interest. The minima of $J_\bq$ are located at the three $M$ points when $J_1$ and $J_2$ are both positive (and hence frustrated) and $1/8 < J_2/J_1 < 1$. 

To understand the nature of the classical ground state manifold when the optimal Luttinger-Tisza modes derive from the wave vectors $\{\bQ_i\}$, consider first the simple case of a Bravais lattice. The triangular and FCC lattices are examples. Up to a global spin rotation, the most general parametrization of the Fourier modes $\bS_{\bQ_i}$ consistent with the fixed length requirement takes the form
\be
\bS_{\bQ_1}= \frac{S}{2}m_x \hat\bx \quad \bS_{\bQ_2}= \frac{S}{2}m_y \hat\by,  \quad
\bS_{\bQ_3}= \frac{S}{2}m_z  \hat\bz, \label{eq:Fourier-S}
\ee
where $S$ is the spin length and $\vec m=(m_x,m_y,m_z)\equiv (\sin\theta\cos\phi, \sin\theta\sin\phi ,  \cos\theta) $ is a unit vector defined in terms of the angles $(\theta,\phi)$. The angles $(\theta,\phi)$ parametrize a family of degenerate classical ground states, which can be represented as points on the Bloch sphere. Generically, such points correspond to triple-$Q$ magnetic states, but there are special lines and points on the Bloch sphere correspond to double-$Q$ and single-$Q$ states. In particular, states which correspond to $\vec m  = (1,0,0)$ and its equivalents define single-$Q$ configurations. 

A natural and convenient way to resolve the degeneracy of the classical ground states is to consider composite order parameters built from the primary magnetic Fourier modes. The distinct symmetry properties of the composite orders unambiguously expose the symmetries of magnetic configurations and can thus be used to sharply distinguish classical ground states. In the present context, we define the nematic composite order parameter $n$ and the chiral order parameter $\chi$ as
\beq
n &=& |\bS_{\bQ_3}|^2 + \omega |\bS_{\bQ_1}|^2 + \omega^2|\bS_{\bQ_2}|^2 , \label{eq:n-def}\\
 \chi &=& \bS_{\bQ_1}\cdot \bS_{\bQ_2} \times  \bS_{\bQ_3}. \label{eq:chi-def}
\eeq
Here $\omega  = e^{2\pi i/3}$ is a cubic root of unity. It is clear from these definitions that both $n$ and $\chi$ transform trivially under translations, but $n$ is time-reversal even whereas $\chi$ is time-reversal odd. Most importantly, since nonzero $n$ must originate from an unequal contribution of the three wave vectors, it signals the breaking of crystal rotation symmetry in the magnetic state. Substituting the Fourier mode representation \eqref{eq:Fourier-S} into \eqref{eq:n-def} and \eqref{eq:chi-def} one finds $n$ and $chi$ in terms of $\vec m$ as
\beq
n & \sim & m_z^2+  \omega m_x^2 + \omega^2m_y^2, \\
 \chi &\sim& m_xm_ym_z.
\eeq
This shows that $n$ is zero when $m_x^2=m_y^2=m_z^2$, and that finite $\chi$ requires all components of $\vec m$ to be nonzero (i.e., noncoplanar order). Note in particular that single-$Q$ ordered states are associated with $\langle n \rangle \neq 0 $. 

Since any energy functional of $n$ contains a cubic Potts-anisotropy $\propto n^3 + (n^*)^{3}$, thus giving rise to three distinct nematic ground states, $n$ describes Pott-nematic order~\cite{}. As a result, $n$ is uniquely associated with the spontaneous breaking of threefold crystal rotation symmetry in the magnetically ordered phase. 

The continuous degeneracy of the classical ground states parametrized by $\vec m$ does not survive the effect of fluctuations. This is due to the well known ``order-by-disorder'' mechanism, which tends to favor collinear spin alignment~\cite{Villain:1980p1263,Henley:1987p3962,Henley:1989p2056,Chandra:1990p88,Jolicoeur:1990p4800,Chubukov:1992p11137}. For instance, whereas the classical energy is independent of the angles $(\theta,\phi)$, the energy associated with the quantum zero-point motion of the magnon excitations does depend on the orientation of $\vec m$, and is lowest for the single-$Q$ states such as $\vec m  = (1,0,0)$~\cite{Jolicoeur:1990p4800,Chubukov:1992p11137,Oguchi:1985p4494,Lecheminant:1995p6647}.

\section{Instabilities of a quantum paramagnet  \label{sec:theory}}

In this section we turn to the central part of this work. As outlined in the introduction, our goal is to determine the instability of a quantum paramagnet towards the formation of nematic magnon bound states. To achieve this, we start from an XXZ model for a system of $S=1$ spins, introduced in Ref.~\onlinecite{Wang:2017p184409}, which is defined by the Hamiltonian
\begin{multline}
\label{eq:H_XXZ}
\mathcal{H} = \frac{1}{2} \sum_{\br, \bdelta} J_{\bdelta} (S_{\br}^{x}S_{\br + \bdelta}^{x} + S_{\br}^{y}S_{\br + \bdelta}^{y} + \eta_{\bdelta}  S_{\br}^{z}S_{\br + \bdelta}^{z}) \\
+ D \sum_{\br}(S_{\br}^{z})^{2}.
\end{multline}
The Hamiltonian is a sum of two terms. The first term is of the general form introduced in \eqref{eq:H_HB} and describes exchange couplings between the spins. Specifically, pairs of spins separated by a distance $\br-\br' = \bdelta$ interact with coupling constant $J_{\bdelta}$. Note that compared to \eqref{eq:H_HB} here we have also introduced an XXZ anisotropy given by $\eta_{\bdelta} \equiv J^z_{\bdelta}/ J_{\bdelta}$, which denotes the ratio of the exchange couplings in the $z$ direction and the $xy$ plane. This is useful, since the parameter $\eta_{\bdelta}$ can be interpreted as controlling the interactions between magnons. As far as the magnetic ground state is concerned, there is no difference with \eqref{eq:H_HB}. The second term in \eqref{eq:H_XXZ} describes a single-ion anisotropy of strength $D>0 $. 

For simplicity, in what follows we focus the discussion on the case of primitive Bravais lattices and leave the straightforward generalization to non-Bravais lattices to Appendix~\ref{app:gen}, the results of which will be used when applying the analysis to the honeycomb lattice. Since we are interested in the quantum behavior of this model we consider the system at $T=0$.  

As argued in Ref.~\onlinecite{Wang:2017p184409}, in the limit where $D$ is much larger than the exchange couplings $J_{\bdelta}$, the ground state is a quantum paramagnet with all spins in the $\ket{0}\equiv\ket{S=1,0}$ state. We denote the paramagnetic ground state as $\ket{\Omega} = \otimes_j \ket{0_j}$. Instead, when $D\rightarrow 0$ one expects a magnetically ordered state, the nature of which is determined by the exchange couplings $J_{\bdelta}$. Crucially, in what follows we assume that the exchange couplings take values such that $J_\bq$ has minima at three degenerate wave vectors $\{\bQ_i\}$, giving rise to a degenerate magnetic ground manifold at the classical level. This is precisely the part of the phase diagram considered in the previous section. In this case, $D$ describes the transition from a quantum paramagnet to a magnetically ordered state which spontaneously breaks a discrete $\mathds{Z}_3$ rotation symmetry. Our goal is to examine this transition and establish whether or not the paramagnetic ground state $\ket{\Omega}$ is unstable towards the formation of \emph{nonmagnetic} two-magnon bound states in the vicinity of the magnetic transition.

\subsection{Schr\"odinger equation for magnon pairs \label{ssec:schroedinger}}
\label{sec:schroedinger_for_magnon_pairs}

To address this question, we follow the approach of Ref.~\onlinecite{Wang:2017p184409} and consider a general two-magnon state $\ket{\psi}$ defined as
\be
\left|\psi\right\rangle =\sum_{\br\neq \br'}\psi_{\br \br'} \left|\br, \br'\right\rangle, \qquad\left|\br, \br'\right\rangle =\frac{1}{2}S_{\br}^{+}S_{\br'}^{-}\left|\Omega\right\rangle , \label{eq:psi-def}
\ee
where $\psi_{\br \br'}$ is the wave function, which is a function of the positions of the two magnons. By projecting Schr\"odinger equation $\mathcal H \ket{\psi} = E\ket{\psi}$ into the two subspace of two-magnon states, we find the equation
\begin{multline}
	\label{Scroedinger_coordinate_space}
	(E - 2D)\psi_{\br \br'} = \\
	\sum_{\bdelta} J_{\bdelta}(\psi_{\br + \bdelta, \br'} + \psi_{\br, \br' + \bdelta}) - J_{\bdelta}^z \, \psi_{\br \br'}  \delta_{\br+ \bdelta, \br' },
\end{multline}
where $J_{\bdelta}^z \equiv \eta_{\bdelta}J_{\bdelta}$. Note that, in accordance with \eqref{eq:psi-def}, the two-magnon wavefunction is understood to vanish when $\br'=\br$ (i.e., $\psi_{\br\br}=0$) \footnote{If one considers ansatz, in which $\ket{\Omega}$ mixes with the two-magnon states, the resulting wavefunction will be dominated by the $\ket{\Omega}$ component as it is lower in energy by at least $2D$ compared to the two-magnon states. The solution can be interpreted as the ground state, corrected by the interaction with the two-magnon states.}.
The first key observation is that the second term on the right hand side, which is proportional to $J_{\bdelta}^z$, generically describes an attractive interaction between the magnons, whereas the first term represents a kinetic term for the two magnons. To proceed, we expand the wave function $\psi_{\br \br'}$ in Fourier modes as
\be
\label{Wigner_coordinates_Fourier}
\psi_{\br \br'}=\sum_{\bK, \bq}e^{i\bK\cdot (\br + \br')/2}e^{i\bq\cdot (\br - \br')} \psi_{\bq}(\bK),
\ee 
where $\bK$ is the momentum conjugate to the center of mass coordinate and $\bq$ is conjugate to the relative coordinate $\br - \br'$. Substituting the Fourier mode expansion into the Schr\"odinger equation \eqref{Scroedinger_coordinate_space} yields the equation
\be
\label{integral_equation}
\left[ E - 2 D - \mathcal E_\bq(\bK) \right]\psi_\bq(\bK) = - \sum_{\bdelta} J_{\bdelta}^z  e^{i \bq \cdot \bdelta} B_{\bdelta},
\ee
where 
\be \label{magnon_pair_energy}
\mathcal E_\bq(\bK) = 2\sum_{\bdelta} J_{\bdelta} e^{i \bK\cdot \bdelta/2} \cos \bq\cdot \bdelta,
\ee
can be viewed as a kinetic energy of the two magnons, and $B_{\bdelta} $, defined as
\be
B_{\bdelta} = \frac{1}{N} \sum_{\bp} e^{ -i \bp\cdot \bdelta} \psi_\bp(\bK), \label{B_def}
\ee
corresponds to the real-space wavefunction as a function of the relative coordinate $\br - \br' = \bdelta$.
Note that in the case of zero center of mass momentum one has $\mathcal E_\bq(\bK=0) \equiv \mathcal E_\bq = 2\varepsilon_\bq$, where $\varepsilon_\bq = \sum_{\bdelta} J_{\bdelta}  \cos \bq\cdot \bdelta$ is the single-magnon dispersion. To solve Eq.~\eqref{integral_equation}, we divide by the kernel $E - 2 D - \mathcal E_\bq(\bK) $ and use the definition of $B_{\bdelta}$ to obtain a matrix equation given by
\be
B_{\bdelta} = \sum_{\bdelta'} \mathcal{M}_{\bdelta \bdelta'} B_{\bdelta'}, \label{matrix_equation}
\ee
where the matrix $\mathcal{M}_{\bdelta \bdelta'}$ is defined as
\be
\label{M_definition}
\mathcal{M}_{\bdelta \bdelta'} = \frac{1}{N} \sum_{\bq} \frac{J_{\bdelta'}^z \ e^{- i \bq\cdot (\bdelta - \bdelta')}}{2D + \mathcal E_\bq(\bK) - E}
\ee
Solutions to the Schr\"odinger equation are then determined by the condition
\be
\text{Det}(\mathcal{M}-\mathbb{1}) = 0.  \label{det_zero}
\ee
In our case, we seek solutions corresponding to zero energy ($E=0$), which means that the energy for creating two-magnon excitations vanishes. The general strategy for solving \eqref{det_zero} after setting $E=0$ is to diagonalize $\mathcal{M}$ and determine when the eigenvalues of $\mathcal{M}-\mathbb{1}$ vanish. This is done as a function of $D$ and the first eigenvalue which vanishes as $D$ decreases corresponds to the solution of interest. 

Inspection of the structure of $\mathcal{M}$ reveals that it is possible to bring it in block diagonal form by making use of the symmetry properties of the underlying lattice and the center of mass momentum $\bK$. To illustrate how group theory machinery can be employed to block diagonalize $\mathcal{M}$, it is useful to consider the example of the triangular lattice
$J_1$--$J_2$ model discussed in the previous section. 

The triangular lattice has six nearest neighbor vectors $\ba_j$ and six next-nearest neighbor vectors $\ba'_j$ \footnote{The six nearest neighbor and next-nearest neighbor vectors are related by sixfold rotation; some are shown in Fig.~\ref{fig:triangular}.}, which we label $j=0,\dots,5$. As a result, $\bdelta,\bdelta'$ in \eqref{matrix_equation} and \eqref{M_definition} take values in this set of twelve vectors. Let us furthermore consider the case $\bK=0$. The matrix $\mathcal M$ is then block diagonalized by a unitary matrix $\mathcal U$ of the form
\be
\mathcal U  = \begin{pmatrix} U &  \\  &U \end{pmatrix},
\ee
where the matrix elements of $U$ are
\be
U_{jl} = e^{i 2\pi l j /6}/\sqrt{6}.
\ee
Here $U$ transforms from the basis of (next)-nearest neighbor vectors to a basis of angular momentum labeled by $l=-2,\ldots,3$. Since the triangular lattice (as well as the center of mass momentum $\bK=\Gamma$) has full hexagonal symmetry, different angular momentum channels cannot mix, implying that $\mathcal U^\dagger  \mathcal M \mathcal U $ is block diagonal. The blocks therefore correspond to distinct angular momentum channels, which we denote $\mathcal{M}_l$ and are given by
\be \label{M_block}
	\mathcal{M}_l=  \frac{1}{N} \sum_{\bq} \frac{1}{2D + \mathcal E_\bq } \begin{pmatrix} J^z_1  | f^l_\bq |^2 & J^z_2 f^{l *}_\bq g^l_\bq \\ J^z_1 g^{l *}_\bq f^l_\bq &J^z_2  | g^l_\bq |^2 \end{pmatrix}. 
\ee
Here $f^l_\bq$ and $g^l_\bq$ are the nearest neighbor and next-nearest neighbor symmetry-adapted lattice harmonics of the triangular lattice and are defined as
\be
	f^l_\bq = \sum_j U_{jl} e^{i \bq \cdot \ba_j}, \quad g^l_\bq = \sum_j U_{jl} e^{i \bq \cdot \ba'_j}. \label{harmonics}
\ee
It is worth noting that the matrix in \eqref{M_block} can be decomposed as
\be
 \begin{pmatrix} J^z_1  | f^l_\bq |^2 & J^z_2 f^{l *}_\bq g^l_\bq \\ J^z_1 g^{l *}_\bq f^l_\bq &J^z_2  | g^l_\bq |^2 \end{pmatrix}=\begin{pmatrix}  f^{l *}_\bq \\ g^{l *}_\bq \end{pmatrix}
	\begin{pmatrix} J^z_1   & 0 \\ 0 &J^z_2  \end{pmatrix} \begin{pmatrix}   f^{l}_\bq &  g^{l}_\bq \end{pmatrix},
\ee
which leads to further simplifications within a weak coupling approach, see Sec.~\ref{ssec:wc}. We further point out that the  $l=0$ angular momentum channel is special and should be excluded, since it leads to a two-magnon wavefunction $\psi_{\br\br'}$ that is nonzero when $\br'=\br$, thus contradicting the assumptions that underlie \eqref{Scroedinger_coordinate_space}.

Equation \eqref{det_zero} now reduces to a set of separate equations in each angular momentum channel given by $\text{Det}(\mathcal{M}_l-\mathbb{1}) = 0$. These equations can be solved---generally numerically---to obtain the critical value $D^*_c$ at which magnon pairs can form with zero energy. Each angular momentum channel will yield a different value of $D^*_c$. The critical values $D^*_c$ are then compared to $D_c$, the value of $D$ at which the gap to single magnon excitations closes, which is obtained by setting $2D_c + \mathcal E_{\bQ_i}=2( D_c + \varepsilon_{\bQ_i})=0$. It thus follows that $D_c$ is determined by the minimum of the magnon dispersion $ \varepsilon_{\bQ_i}$. (Note that $\varepsilon_{\bQ_i}$ is negative.) When $D^*_c > D_c$ for one or more $l$, two-magnon bound states can form with positive binding energy
\be
\varepsilon_b = 2(D^*_c -D_c). \label{eq:eps_b-def}
\ee
The angular momentum channel with the largest value $D^*_c$ then determines the structure of excitons, much like the structure of Cooper pairs is determined by the relative angular momentum of the constituent electron pairs.

It is worth noting that symmetries generally imply degeneracies of angular momentum channels. For instance, in the case of the triangular lattice, $l=\pm 2$ (as well as $l=\pm 1$) are degenerate and form a single irreducible channel.

This detailed example of the triangular lattice serves to illustrate the general symmetry-based method for solving \eqref{det_zero}. 
In particular, the matrix $\mathcal M$ can always be block diagonalized such that each block $\mathcal{M}_i$ corresponds to distinct symmetry quantum numbers (i.e., representation of the lattice symmetry group) and matrix elements within each block are products of symmetry-adapted lattice harmonics. 
So, Eq.~\eqref{det_zero} reduces to a set of decoupled equations in each symmetry channel of the form $\text{Det}(\mathcal{M}_i - \mathbb{1}) = 0$.
Their solutions  determine whether two-magnon bound states can form and if so, in what symmetry channel they form. 
This general method applies to finite center-of-mass momentum $\bK$ as well, with the modification that the relevant symmetry group is the little group of $\bK$.

\subsection{Weak coupling \label{ssec:wc}}

Further analysis of Eq.~\eqref{M_block} is possible when the interaction between the magnons is small. It follows from Eq.~\eqref{Scroedinger_coordinate_space} that the interaction between magnons originates from the $S^z S^z$ coupling and is parametrized by $\eta_{\bdelta} = J_{\bdelta}^z / J_{\bdelta}$, which may be treated as a perturbative parameter. Similar to Ref.~\onlinecite{Wang:2017p184409}, this assumption will allow us to derive a compact expression for the exciton binding energy, which is shown to depend on: (i) an effective coupling constant and (ii) magnon density of states.

In Section~\ref{sec:schroedinger_for_magnon_pairs}, we used \eqref{Wigner_coordinates_Fourier} to express the two-magnon problem to Schr{\"o}dinger equation in terms of center-of-mass and relative coordinates. Under this transformation, the magnon interaction translates into quantum well potential for the relative coordinate $\br - \br'\ = \bdelta$.  As a result, the weak coupling limit is equivalent to the shallow well approximation in quantum mechanics. This is why we expect the bound state binding energy $\varepsilon_b$ to be small compared to the exchange couplings $J_{\bdelta}$, while the size of the two-magnon bound state (i.e., the ``exciton'') should be large on the lattice scale. Consequently, the extent of the exciton in reciprocal space will be small, which allows us to expand the dispersion \eqref{magnon_pair_energy} in the vicinity of its minima. 

\begin{figure*}
	\includegraphics[width=\textwidth]{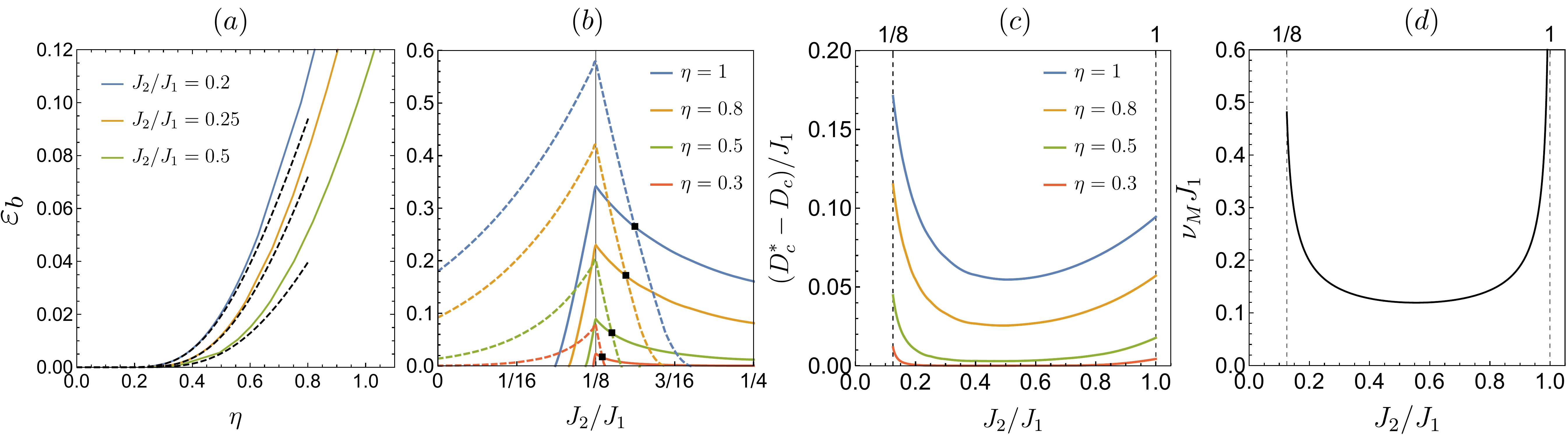}
	\caption{Two-magnon ``exciton'' bound states in the triangular lattice model. 
	(a) Binding energy $\varepsilon_b$ defined in Eq.~\eqref{eq:eps_b-def} as a function of $\eta = J^z_{1,2}/J_{1,2}$ for different values of $J_2/J_1$. Solid (and colored) curves correspond to numerical solutions of Eq.~\eqref{M_block} for $l=2$. Dashed (and black) curves correspond to the binding energy computed using Eq.~\eqref{binding_energy} (with prefactor fitted), showing excellent agreement for small $\eta$.  
	(b) The binding energies as a function of $J_2/J_1$ computed numerically from Eq.~\eqref{M_block} for the $l=2$ (solid curves) and $l=3$ (dashed curves) angular momentum channels. Black squares indicate where the nematic $l=2$ channel becomes dominant.
	(c) The binding energies as a function of $J_2 / J_1$ in the interval $1/8 < J_2 / J_1 < 1$ (bounded by vertical dashed lines).
	(d) Single-magnon density of states $\nu_M$ (in units of $J^{-1}_1$) for magnon excitations on top of the paramagnetic phase evaluated at the $M$ points. $\nu_M$ is given by Eq.~\eqref{DOS_triangular} and explains the general behavior of $\varepsilon_b$ shown in (c).
	}
	\label{fig:triangular_results}
\end{figure*}

To illustrate the weak coupling analysis, we will again consider the case of triangular lattice, and relegate the generalization to non-Bravais lattices to Appendix~\ref{app:gen}. Following the scheme outlined above, we assume that the summand in \eqref{M_block} is concentrated in the vicinities of the wave vectors $\bQ_i$, the locations of the minima of \eqref{magnon_pair_energy}.  Assuming further that $\bK=0$, we expand $\mathcal{E}_\bq$ in the denominator up to the second order in $\bp = \bq - \bQ_i$ and evaluate the lattice harmonics in the numerator at $\bQ_i$. We also use the fact that the summand is invariant under rotation of $\bq$ by $2\pi / 3$, which allows to consider just one out of three $\bQ_i$ points; the contributions from each wave vector must be equal. 
Then the matrix structure of \eqref{M_block} is determined by a single $\bq$ point and has rank one. Evidently, for fixed angular momentum $l$ it has just one nonzero eigenvalue corresponding to the eigenvector $(  f^{l}_{\bQ_i} , g^{l}_{\bQ_i})^T$. That leads to the following equation, where the summation was replaced by integration in the vicinity of $\bQ_i$:
\be  \label{final_eq_expand}
	  \lambda^l \int \frac{d\bp}{V_{\text{BZ}}} \frac{1}{\bp^T A_i\bp +\varepsilon_b} = 1.
\ee
Here $V_{\text{BZ}}$ is the area of the Brillouin zone, $\bp$ is a small momentum reckoned from $\bQ_i$, and $(A_i)_{mn} = \partial^2 \mathcal E_\bq / \partial q_m\partial q_n |_{\bQ_i} $. The effective coupling constants $\lambda^l$ and the exciton binding energy $\varepsilon_b$ are defined as 
\be \label{lambda_general}
	\lambda^l = 3 \big( J^z_1  | f^l_{\bQ_i} |^2+ J^z_2  | g^l_{\bQ_i} |^2 \big), \quad \varepsilon_b = 2D + \mathcal E_{\bQ_i}. 
\ee
Integral \eqref{final_eq_expand} is most easily performed by switching to the integration over single-magnon energy $\xi = \bp^T A_i\bp / 2$:
\be \label{final_eq_log}
	\nu_{\bQ_i} \lambda^l  \int d \xi \frac{1}{2 \xi +\varepsilon_b} =  \frac{\nu_{\bQ_i}  \lambda^l}{2} \log\frac{\omega_c}{\epsilon_b}  = 1, 
\ee
where $\omega_c$ is a cutoff on the order of the exchange energies $J_{1,2}$ and $\nu_{\bQ_i}$ is magnon density of states at one of the three minima.
The latter is finite in two dimensions at a quadratic dispersion minimum and can be evaluated by $A_i$ diagonalization  followed by a linear coordinate transform for the integration variables $p_x, p_y$. 
If $\rho_1$ and $\rho_2$ are the eigenvalues of $A_i$ (obviously, they are the same for each $\bQ_i$), then $\nu_{\bQ_i} \propto 1/\sqrt{\rho_1\rho_2}$.
Note that flattening of the minima enhances the density of states and increases the binding energy.
The value of the latter follows from \eqref{final_eq_log}:
\be \label{binding_energy}
	\epsilon_b = \omega_c \exp\left(- \frac{2}{ \nu_{\bQ_i} \lambda^l}\right), 
\ee
which is a key result of this section.  The expression for the binding energy closely resembles the one for the Cooper problem of two attracting electrons above the two-dimensional Fermi surface. That is expected as both problems are in a weak coupling regime and effectively two spacial dimensions. 

The energy of the two-magnon state is given by the familiar expression for two-particle bound states, i.e., $E = 2(D + \varepsilon_{\bQ_i}) - \varepsilon_b$, which implies that zero energy solutions exist for a critical value $D^*_c = -\varepsilon_{\bQ_i} + \varepsilon_b/2 =  D_c+ \varepsilon_b/2$, reproducing Eq.~\eqref{eq:eps_b-def}. Since the binding energy is positive for channels with nonzero (attractive) coupling constant, one has $D^*_c > D_c$, demonstrating that the quantum paramagnet is unstable towards the formation of non-magnetic two-magnon excitons. The internal structure of the excitons, as defined by their symmetry quantum numbers, directly follows from comparing the coupling constants $\lambda^l$: the angular momentum channel with the largest coupling gives rise to the largest value of $D^*_c$ and determines the leading instability. As explained in the previous Section, zero angular momentum channel should be eliminated.

We expect the result $\eqref{binding_energy}$ to be general for two-dimensional materials with three equivalent minima of the magnon dispersion. 
Details of the model in each case are encoded into magnon density of states and effective coupling constants $\lambda^l$. 
For non-Bravais lattices, Eq.~\eqref{lambda_general} should be modified by introducing projectors to the lower magnon band as explained in Appendix~\ref{app:gen}.
We also comment on generalization to three spacial dimensions in the next Section.


\section{Potts-nematic magnon bound states \label{sec:results}}

In this section, which forms the heart of the paper, we present a detailed application of the theory developed in the previous section to three particular lattice models that exhibit rotation symmetry broken magnetism. Specifically, we consider the triangular and honeycomb lattice in two dimensions, and the FCC lattice in three dimensions.  We present both the outcome of the exact solution of the two-particle problem obtained by numerical integration in Eq.~\eqref{M_block} and the prediction of the analytical perturbation theory given by \eqref{final_eq_log}. For the case of the FCC lattice we show how weak coupling approach, which in two dimensions gives rise to \eqref{binding_energy}, still yields useful approximations to the numerically exact result.

\subsection{Triangular lattice magnets \label{ssec:triangular}}

We begin by considering the triangular lattice model briefly introduced in Sec.~\ref{sec:Z3_breaking} (see also Fig.~\ref{fig:triangular}). This model includes antiferromagnetic exchange couplings between nearest neighbors, given by $J_1>0$, and next-nearest neighbors, given by $J_2>0$, which define the XXZ part of \eqref{eq:H_XXZ}. Since both couplings are antiferromagnetic the interactions are frustrated. We specifically focus on the part of the phase diagram where $1/8 < J_2 / J_1 < 1$, since in this regime $J_{\bq}$ has minima at three commensurate $M$-point wave vectors (i.e., the centers of the edges of the Brillouin zone), thus implying a magnetically ordered state of collinear stripes at small $D$ \cite{Szasz:2022p115103} [shown in Fig.~\ref{fig:triangular}(b)]. Furthermore, since the dispersion of the single-magnon excitations in the quantum paramagnetic state is proportional to $J_{\bq}$ (i.e., $\varepsilon_{\bq}\sim J_\bq$, see Sec.~\ref{sec:schroedinger_for_magnon_pairs}), its minima are also located at the $M$ points.

\begin{figure}
	\includegraphics[width=0.8\columnwidth]{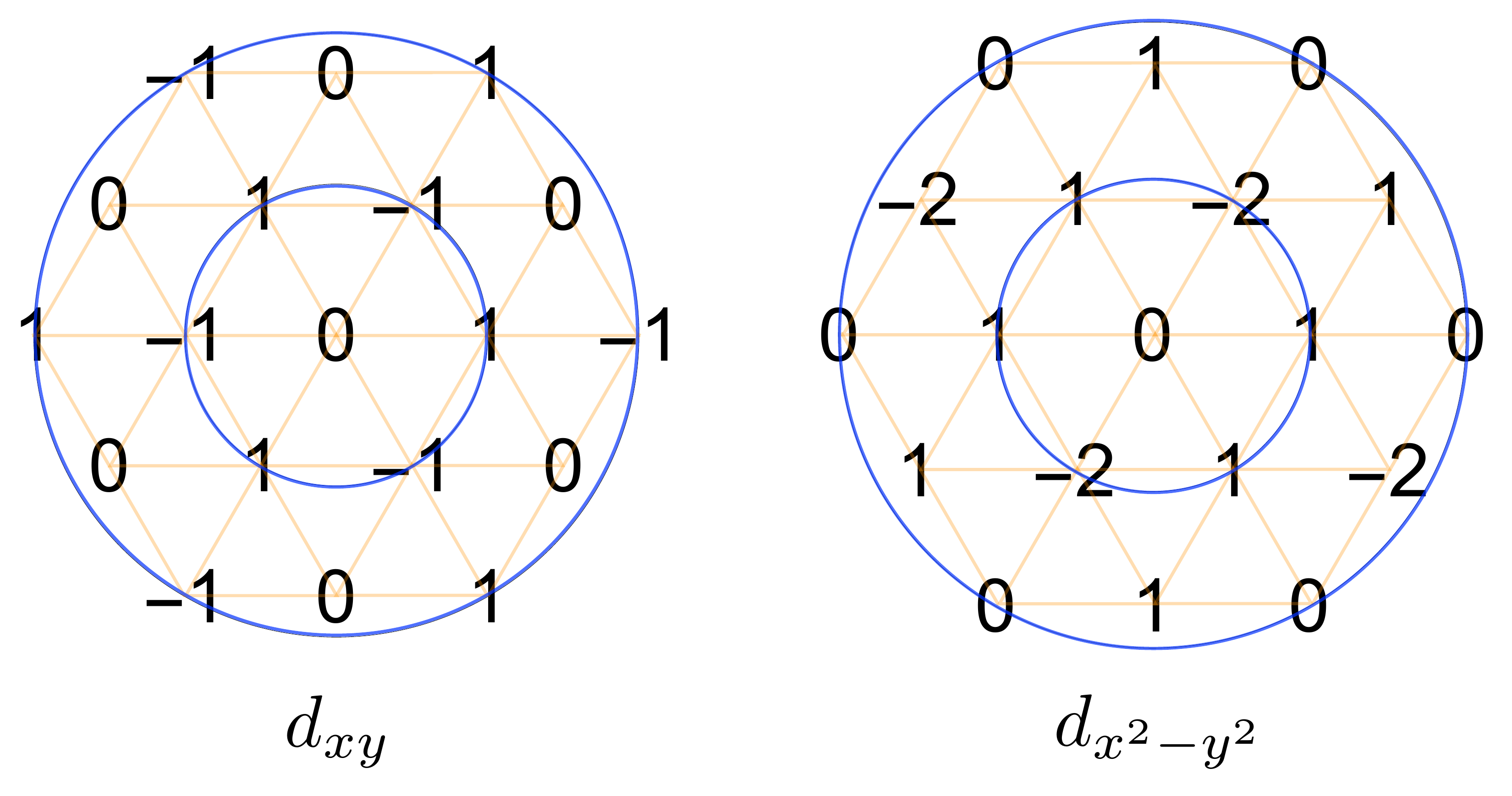}
	\caption{Real space structure of the two-magnon bound state wave function on the triangular lattice in the $l=2$ angular momentum channel. Shown are the wave functions $\psi_{1,2}(\brho) $ given in Eq.~\eqref{eq:wf_triangular}, where $\brho = \br-\br'$ is the distance between the magnons.
	Doubly degenerate solution for the exciton wavefunction on the triangular lattice in the weak coupling regime, shown in real space [$\psi(\brho) = (1/2\pi) \int d \bq\, \psi_\bq e^{i \bq\cdot  \brho} $, where $\brho$ is the vector connecting the two magnons].
	The values slowly decay in space due to the finite exciton size (not shown).
	}
	\label{fig:triangular_wavefunction}
\end{figure}

Consider first the numerical solutions of Eq.~\eqref{M_block}. These are obtained straightforwardly by numerical evaluation of the sum and the key results are presented in Fig.~\ref{fig:triangular_results}. Most importantly, we find that in the regime of interest, i.e., $1/8 \lesssim J_2 / J_1 < 1$, the binding energy for two-magnon bound states is positive and largest in the $l=2$ angular momentum channel, which is the doubly degenerate (Potts-)nematic channel. In Fig.~\ref{fig:triangular_results}(a) we show the binding energy $\varepsilon_b$ obtained from solving \eqref{M_block} for $l=2$ as a function of $\eta$, which controls the strength of the attractive magnon interaction. For convenience here we have taken $\eta_{1,2} = \eta$, where $\eta_i = J^z_{i} / J_i $, i.e., the ratio between $J^z$ and $J$ is equal for nearest and next-nearest neighbor couplings. Different (solid and colored) curves correspond to different values of $J_2/J_1$, showing that the binding energy increases as $J_2/J_1$ decreases. For comparison, the dashed black curves show the binding energy calculated using Eq.~\eqref{binding_energy}, which was derived under the assumption that $\eta$ is small. We see that the BCS-like formula \eqref{binding_energy} is in excellent agreement with the numerically exact result in this regime. 

In panels (b) and (c) of Fig.~\ref{fig:triangular_results} we show the binding energy, or equivalently $D^*_c-D_c$, as a function of the strength of $J_2$ relative to $J_1$ for various values of $\eta$. As expected, panel (c) shows that the binding energy increases for increasing $\eta$. Furthermore, we find that the binding energy is a convex function of $J_2/J_1$ and increases towards to boundaries of the interval $1/8 < J_2 / J_1 < 1$. As discussed below, this behavior can be explained by \eqref{binding_energy}, in particular the density of states $\nu_M$ at the $M$ points.

The exchange coupling ratio $J_2/J_1 = 1/8$ is of special importance since the minima of $J_\bq$ change at this value, thus implying that the classical ground state changes. In particular, for $J_2/J_1 < 1/8$ the minima of $J_\bq$ are located at the $K$ points, giving rise to the well-known coplanar $120^\circ$ spiral state on the classical level. Note that in the vicinity of $J_2/J_1 = 1/8$ both the $M$ points and the $K$ are local minima of $J_\bq$; it is the global minimum which changes at $J_2/J_1 = 1/8$. 

In the limit $J_2/J_1 = 0 $, Ref.~\onlinecite{Wang:2017p184409} showed that two-magnon bound states can form in the (nondegenerate) $l=3$ channel, which is the channel naturally associated with the $K$ points. Excitons with $l=3$ break an Ising $\mathds{Z}_2$ symmetry, which corresponds to the chirality of the $120^\circ$ classical order. To examine the behavior in the vicinity of $J_2/J_1 = 1/8$ we compute the binding energy in the interval $0 < J_2/J_1 < 1/4$ for both channels and different values of $\eta$, and show the result in panel (b) of Fig.~\ref{fig:triangular_results}. As is clear, in the close vicinity of $J_2/J_1 = 1/8$ the binding energy is positive in both channels, but is larger for the ``chiral'' $l=3$ channel. At some value $J_2/J_1 \gtrsim 1/8$, which is marked by a black squares in Fig.~\ref{fig:triangular_results}(b) and increases with $\eta$, the ``nematic'' $l=2$ channel becomes dominant, implying a transition from a chiral instability towards a Potts-nematic instability. Note that the curves exhibit a cusp (i.e., discontinuous derivative) at  $J_2/J_1 = 1/8$, which is due to the aforementioned abrupt change of the global minimum. 

\begin{figure}
	\label{fig:nonzero_CoM}
	\includegraphics[width=0.8\columnwidth]{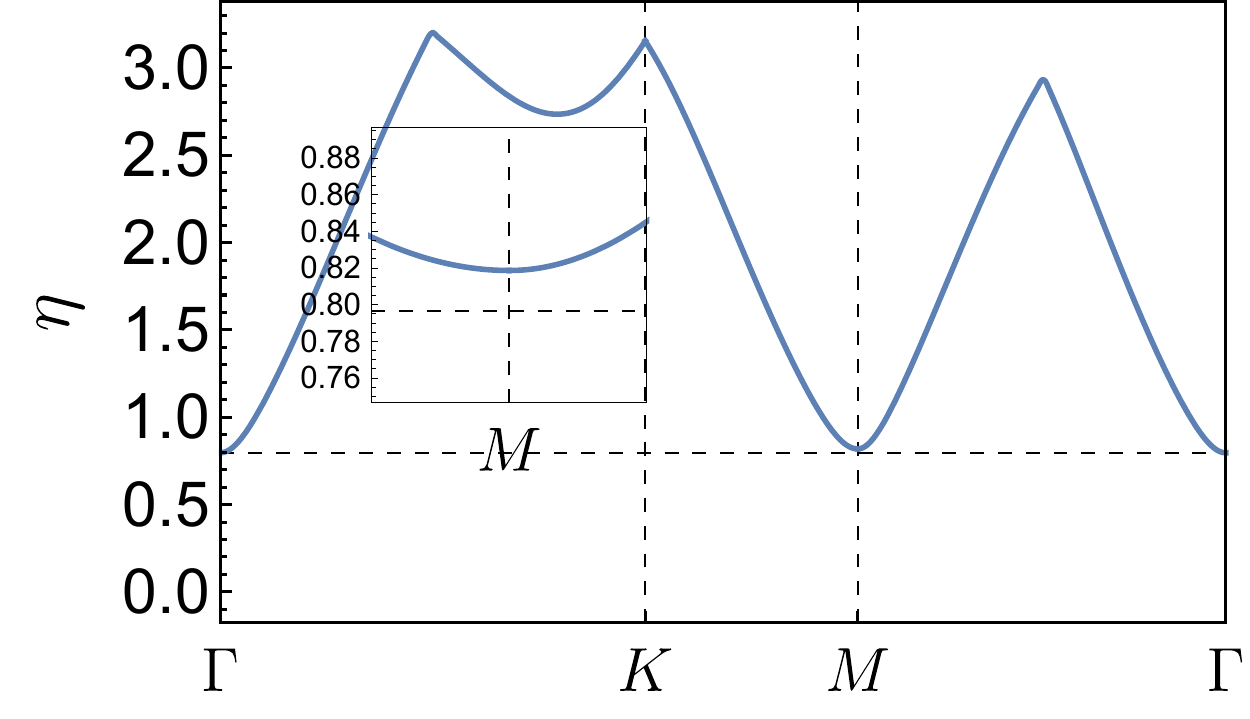}
	\caption{The value of the  interaction parameter $\eta = J_i^z / J_i$ required to create an exciton with the binding energy $\varepsilon_\bq = 0.05 J_1$ for different values of the center-of-mass momentum $\bK$, taking values along the $\Gamma$-$K$-$M$-$\Gamma$ contour in the Brillouin zone (horizontal axis). The minimal $\eta$ corresponds to the $\bK = 0$ case ($\Gamma$ point), which makes it the leading instability channel and justifies considering zero center-of-mass momentum excitons in the rest of the paper.
}
\end{figure}

To further understand the formation of nematic two-magnon bound states, we follow the weak coupling approach described in Sec.~\ref{ssec:wc} and compute the magnon density of states $\nu_M^\triangle $ and effective coupling constant $\lambda_\triangle$ by expanding around the $M$ points. The density of states is given by 
\be
\label{DOS_triangular}
	\nu_M^\triangle = \frac{1}{4 \pi \sqrt{(9J_2 - J_1)(J_1 - J_2)}}  
\ee
and the effective coupling constant is found as
\be
\lambda_\triangle = 8(J_1^z + J_2^z). 	\label{lambda_triangular}
\ee
These two parameters enter the expression for the binding energy in Eq.~\eqref{binding_energy} and have been used to compute the black dashed curves in Fig.~\ref{fig:triangular_results}(a). (The prefactor coming from the cutoff was fitted.) The density of states, which is shown in panel (d) of Fig.~\ref{fig:triangular_results}, is singular and diverges at $J_2=J_1$. This is caused by the fact that for $J_2>J_1$ the minima of $J_\bq$ move away from $M$ towards $\Gamma$, making the $M$ points saddle points rather than minima. In contrast, $\nu_M^\triangle$ is regular at $J_2/J_1=1/8$, because $M$ remains a local minimum.  Since the effective coupling \eqref{lambda_triangular} remains finite and does not change significantly in the interval $1/8 < J_2/J_1 < 1$, the main dependence of  the binding energy $\varepsilon_b$ on $J_2 / J_1$ ratio comes from $J_2 / J_1$ dependence of $\nu_M$, which is illustrated by Figs.~\ref{fig:triangular_results}(c) and (d).

\begin{figure*}
	\includegraphics[width=\textwidth]{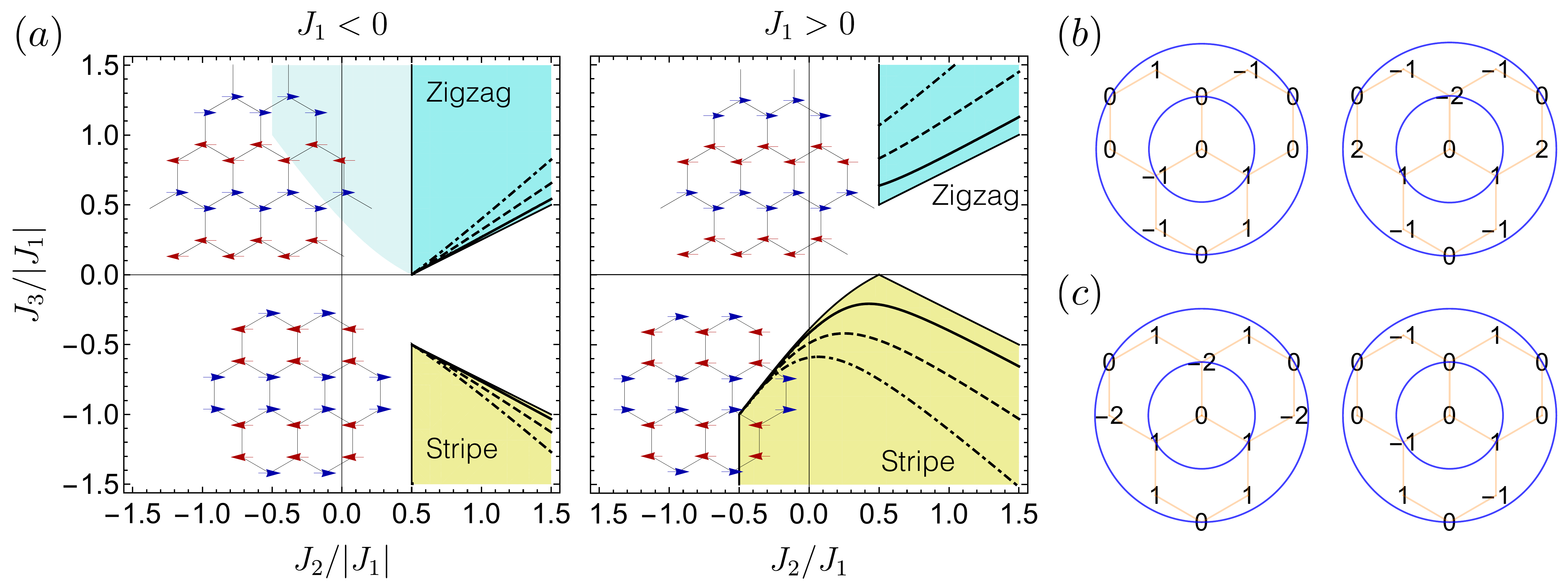}
	\caption{
	(a-1) Zigzag (blue) and striped (yellow) phases on the honeycomb lattice for antiferromagnetic nearest neighbor coupling ($J_1 < 0 $ case). Effective Potts-nematic coupling constant \eqref{lambda_honeycomb} is positive only at $J_2 > J_1/2$, so vestigial state can emerge only in the part of the zigzag domain (shown in darker blue).
Lines demonstrate contours at which perturbative theory predicts the width of the intermediate phase to be $(D_c^* - D_c) / J_1 \sim 10^{-1}$ (solid), $\sim 10^{-2}$ (dashed), and $ \sim 10^{-3}$ (dot-dashed). We assumed $J_i^z = J_i$ when plotting this picture.
(a-2) 
Similar plot for $J_1 > 0$ case. In that case vestigial phase can appear for all points inside the zigzag and striped magnetic domains.
(b) (c) 
The core real-space ($\brho$) structure of the Potts-nematic exciton on the honeycomb lattice in the perturbative regime for a) zigzag phase; b) striped phase. The spin-up magnon resides on the first sublattice, while the spin-down magnon is located at a site, shifted by $\brho$. The solution is doubly degenerate in each case. Slow exponential decay is not shown. 
	}
	\label{fig:honeycomb_results}
\end{figure*}

As explained in Section~\ref{sec:schroedinger_for_magnon_pairs}, excitonic solution in the Potts-nematic channel correspond to $l=\pm2$ values of the angular momentum and thus has a twofold degeneracy, which we observe both in numerical and weak coupling approaches. 
In the latter, it is straightforward to obtain an explicit expression for the wavefunctions in the real space, which up to a constant factor reads:
\be
	\psi_{1(2)}(\brho) = \text{Re(Im)}\left[ e^{i \bM_1\cdot \brho} + \omega e^{i \bM_2 \cdot\brho} + \omega^2 e^{i \bM_3 \cdot\brho} \right] e^{-r / l_\varepsilon}, \label{eq:wf_triangular}
\ee
Here $\brho$ is the vector connecting the two lattice sites occupied by magnons constituting the exciton and $\omega = e^{2\pi i /3}$ is the cubic root of unity.
The factor $e^{-r / l_\varepsilon}$, where $l_\varepsilon \sim \varepsilon^{-1/2} \gg 1$ appears in the next orders of the perturbation theory and makes the wavefunction normalizable. 
Eq.~\eqref{eq:wf_triangular} represents the fact that the solution is a linear combination of exponentials $e^{i \bM_i \cdot\brho}$ with the wavevectors $\bM_i$ pointing to the minima of the single-magnon dispersion, so that the exciton is formed form the most low-lying excitations. The three exponentials are combined in a function with $l=\pm2$ in the same way as the nematic order parameter \eqref{eq:n-def} is constituted from the spin densities at $\bM_i$ points. Taking then real (imaginary) part of $l=2$ wavefunction [Eq.~\eqref{eq:wf_triangular}] gives the wavefunctions with the symmetry of $d_{xy}$ ($d_{x^2 - y^2}$) orbitals.
The third linearly independent combination of exponentials $e^{i \bM_1\cdot \brho} + e^{i \bM_2 \cdot\brho} + e^{i \bM_3\cdot \brho}$ provides the rotation-symmetric $l=0$ wavefunction, which contradicts assumptions made in \eqref{Scroedinger_coordinate_space} and thus should be discarded.

We show the bound state wavefunctions \eqref{eq:wf_triangular} graphically in Fig.~\ref{fig:triangular_wavefunction} ignoring the slow decaying factor. 
Note that the radial structure is not completely trivial: one can see that the angular harmonics are not aligned on the neighboring shells but are rotated with respect to each other.
That makes $\psi(\brho)$ obey the approximate translational symmetry $\brho \rightarrow \brho + 2 \bdelta_{1(2)}$ with doubled period reflecting the fact the $2 \bM_i \sim 0$ in the reciprocal space.
(The translation symmetry is not exact due to the decaying factor $e^{-\rho / l_\varepsilon}$.)
Note also that the excitonic wavefunction $\psi(\brho)$ nullifies when the two magnons reside on the third-nearest sites.
In accordance with this, we found that introducing third-nearest neighbor interaction $J_3$ does not change the value of the effective coupling \eqref{lambda_triangular}.

Finally, to make sure that excitons with zero center-of-mass momentum $\bK$ represent the leading instability, we run a numerical calculation for the case of a non-zero $\bK$, which still boils down to solving \eqref{det_zero} with $\mathcal M$ given by Eq.~\eqref{M_definition}.
In this place we opted for fixing the binding energy at the value of $\varepsilon = 0.05 J_1$ and evaluating the minimal interaction strength $\eta$ required to create such bound state.
The result is presented in Fig.~\ref{fig:nonzero_CoM}, where we allowed the center-of-mass momentum $K$ to take values across a path in the Brillouin zone, connecting the points $\Gamma-K-M-\Gamma$.
Indeed, we observed that $\bK = 0$ requires the minimal amount of interaction for the bound state formation, which justifies considering $\bK=0$ in the rest of the paper.
Interestingly, we found that $\bK = \bM_i$ choice is also favorable for exciton formation, however, this instability remains subleading.

\subsection{Honeycomb lattice magnets \label{ssec:honeycomb}}

Next, we examine Potts-nematic two-magnon bound state formation on the honeycomb lattice. The motivation for considering the honeycomb lattice is twofold. First, it has the same symmetry group as the triangular lattice, but contrary to the latter it has sublattice structure, thus providing a generalization of the analysis presented in Sec.~\ref{sec:theory} to non-Bravais lattices. Second, a variety of experimentally available candidate materials are realizations of the honeycomb lattice model which makes the honeycomb lattice highly relevant from an experimental perspective. The connection to experimental compounds will be discussed in more detail below in Sec.~\ref{sec:expts}.

As mentioned in Sec.~\ref{sec:Z3_breaking}, as far as the exchange interactions included in Eq.~\eqref{eq:H_XXZ} are concerned, we consider a frustrated $J_1$--$J_2$--$J_3$ honeycomb model with up to third-nearest neighbor couplings. The classical phase diagram of the isotropic $J_1$--$J_2$--$J_3$ Heisenberg model is known~\cite{Fouet:2001p241, Gong:2015p195110} and notably includes two magnetic phases with broken threefold rotation symmetry. The corresponding regions in the phase diagram are defined by property that the minima of $J_\bq$ are located at the $M$ points. Note that since $J_\bq$ is a matrix in the space of the two sublattices, two distinct solutions exist, which are related by the relative orientation of the spins on the sublattices, i.e., either aligned or anti-aligned. The single-$Q$ collinear orderings with $M$-point wave vectors that are selected by fluctuations are schematically shown in Fig.~\ref{fig:honeycomb_results}, and are generally referred to as stripe (sublattice-even) and zigzag (sublattice-odd) orders. In our analysis we restrict to exchange couplings ($J_1,J_2,J_3$) which favor zigzag or stripe order on the classical level.

\begin{figure}
	\includegraphics[width=0.98\linewidth]{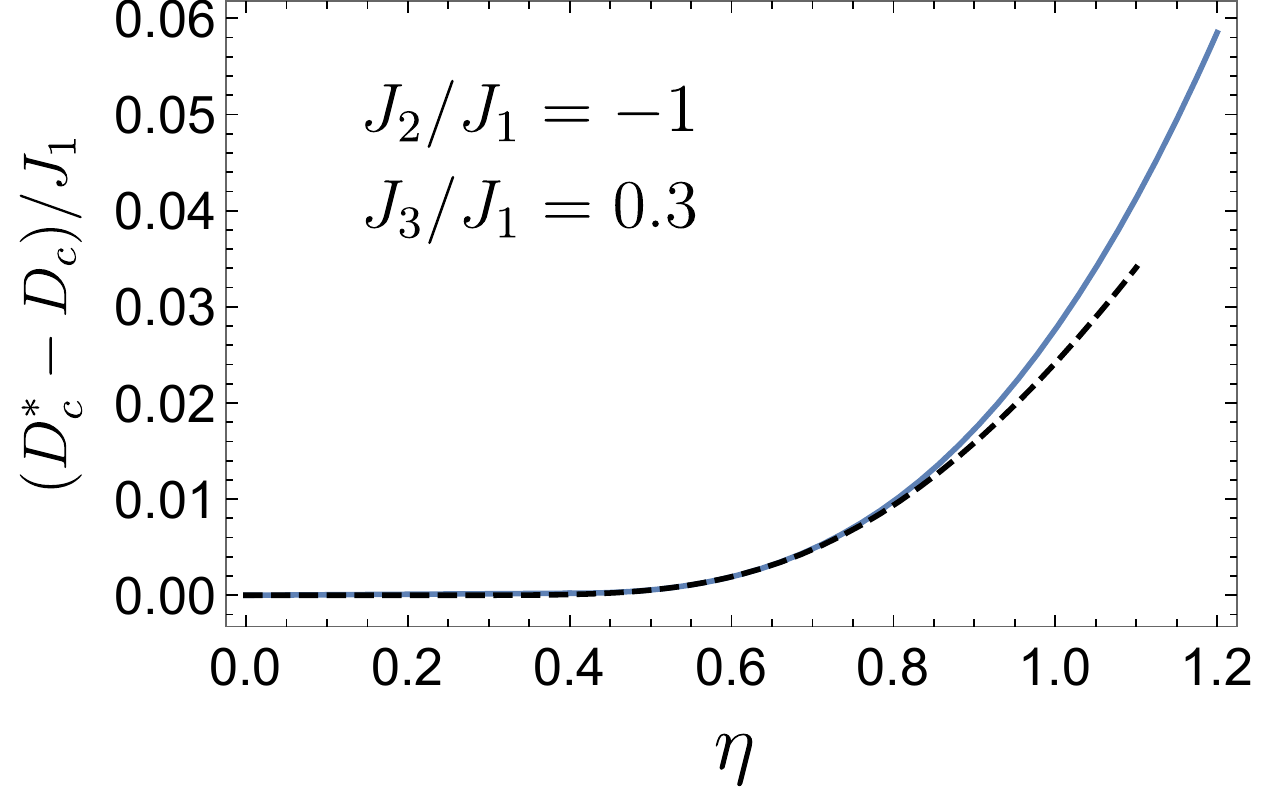}
	\caption{
	The width of the Potts-nematic phase $(D_c^* - D_c)/J_1$ obtained from exciton binding energy as a function of interaction strength $\eta  = J_i^z / J_i$ for the honeycomb lattice model. Solid line shows numerical solution, while the dashed one is for the analytical perturbative result, valid at $\eta \ll 1$. The prefactor in the latter was fitted. }
\label{fig:honeycomb_comparison}
\end{figure}

To account for the two sublattices in the system, we extend the general theory presented in Sec.~\ref{sec:schroedinger_for_magnon_pairs} to the case of a non-Bravais lattice (see Appendix~\ref{app:gen}).
The main occurring difference is the appearance of projectors to the Bloch bands in the expression for the effective matrix elements \eqref{M_multiband}.
We present solutions of the modified equations obtained by numerical Brillouine zone integration complemented by the band summation in Fig.~\ref{fig:honeycomb_comparison}. For that plot, we picked values of the spin couplings inside the zigzag domain on the Fig.~\ref{fig:honeycomb_results}~(a) ($J_1 < 0, J_2 / |J_1| = 1, J_3 / |J_1| = 0.3$) and varied the effective interaction strength $\eta = \eta_i = J_i^z / J_i$.
We observed that bound state with the angular momentum $l=\pm2$ has the largest binding energy and its dependence on $\eta$ is captured well by the perturbation theory with a fitted prefactor (dashed line).
As in the case of triangular lattice, that means that proliferation of excitations of this kind constitutes the leading instability, when $D$ is lowered and that marks the transition to the Potts-nematic phase.

The weak coupling theory for a non-Bravais lattice in general follows the previous derivation given in Section~\ref{ssec:wc}. Again, the main contribution stems from the most low-lying excitations, which now also means that it is sufficient to consider the lower band. We expand around the minima of the latter (located at $\bM_i$ points) and, after some algebra, get expressions for the effective coupling constant in the Potts-nematic channel 
\be
\label{lambda_honeycomb}
	\lambda_{\hexagon} = 8 ( J_1^z + 2 J_2^z ),
\ee
which appears to be the same for both zigzag and stripe $M$-orderings. The density of low-lying states at $M$ points equals:
\be
\label{dos_honeycomb}
	\nu_M^{\hexagon} = \frac{\sqrt{|J_1 - 3 J_3|} / (2\pi J_1)}{\sqrt{(J_{13} - 2 J_2 |J_1 - 3 J_3|)(2J_2 + \beta (4 J_3 - J_1))}},
\ee
where $J_{13} = J_1^2 + J_1 J_3 -4 J_3^2$ and $\beta = \text{sign}\,(J_1 - 3 J_3)$. 
Eqs.~\eqref{lambda_honeycomb} and \eqref{dos_honeycomb} should be plugged into the general perturbative formula for the binding energy \eqref{binding_energy}, which gives a result consistent with numerical computation (Fig.~\ref{fig:honeycomb_comparison}).

Perturbative wavefunction of $l=\pm2$ excitions are again doubly degenerate and appear to be a linear combination of the exponentials $e^{i \bM_i  \br}$ as in the case of triangular lattice, but bear a sublattice structure, which slightly differs between the zigzag and stripe phase. We present it graphically in Fig.~\ref{fig:honeycomb_results}~(b) for underlying zigzag ordering an Fig.~\ref{fig:honeycomb_results}~(c) for the stripe one, choosing the  $d_{xy}$ and $d_{x^2 - y^2}$ basis and using the $\brho$ plane, where $\brho$ is the distance between the magnons in the pair. We assumed that the spin-up magnon resides on the first sublattice; the opposite case leads to similar pictures. 
The figure depicts the core of the exciton wavefunction with omitted slow decay factor $e^{-\rho / l_\varepsilon}$.

With obtained analytical approximation, we explore exciton formation for all striped phases in the phase diagram of Fig.~\ref{fig:honeycomb_results}~(a).
The fist important observation is that for the bound state to emerge, the effective coupling constant \eqref{lambda_honeycomb} needs to be positive. 
It turns out that at $J_1 < 0$ and inside the zigzag domain ($J_3>0$), $\lambda_{\hexagon} > 0$ only when $J_2 / J_1 > 1/2$, which defines a subregion potent to develop a vestigial phase (shown by darker tone and solid frame). Contrary to that, for the stripe phase ($J_3<0$) and for both phases at $J_1 > 0$, the effective coupling is positive everywhere inside the magnetic domain, so that in the latter cases magnetic ordering should always be preempted by a vestigial phase.

We now turn to the discussion of how prominent should be the effect of exciton formation across the honeycomb lattice phase diagram.
As suggested by Eq.~\eqref{binding_energy} and common knowledge, enhancement of the density of states at minima of the dispersion is beneficial for the formation of the bound state. 
Inspection of \eqref{dos_honeycomb} implies that such enhancement happens in the vicinity of the right boundaries of zigzag and stripe domains.
That can be explained by the fact that at these lines the system undergoes phase transition into the spiral ordering, characterized by a non-commensurate value of ordering wavevector $\bQ$.
After the transition, the minima of the coupling matrix $J_{\bq}$ eigenvalues and, consequently, the minima of the single-magnon dispersion on top of paramagnet drift away from the $\bM$-points, turning the $\bM$-point itself into a saddle point rather than the minimum. Before the transition but near the boundary the effect is manifested by appearing of the soft direction at each $\bM$-point, leading to the increase of the single-magnon density of states.
This phenomenon is completely analogous to the one, which happens on the triangular lattice in the vicinity of $J_2/J_1 = 1$ point as described in Section~\ref{ssec:triangular}.
To further illustrate this statement and make it quantitative, we plot contours, at which our perturbative theory predicts the binding energy (and hence the width of the Potts-nematic phase) to be of the order $10^{-1}$ (solid line), $10^{-2}$ (dashed line), and $10^{-3}$ (dot-dashed line) in the units of $|J_1|$ [Fig.~\ref{fig:honeycomb_results}~(a)]. 
As can be seen in the picture, the most perceptible effect appears near the aforementioned boundary of the domains.

\black
\subsection{FCC lattice magnets \label{ssec:fcc}}

\begin{figure}
	\includegraphics[width=0.95\linewidth]{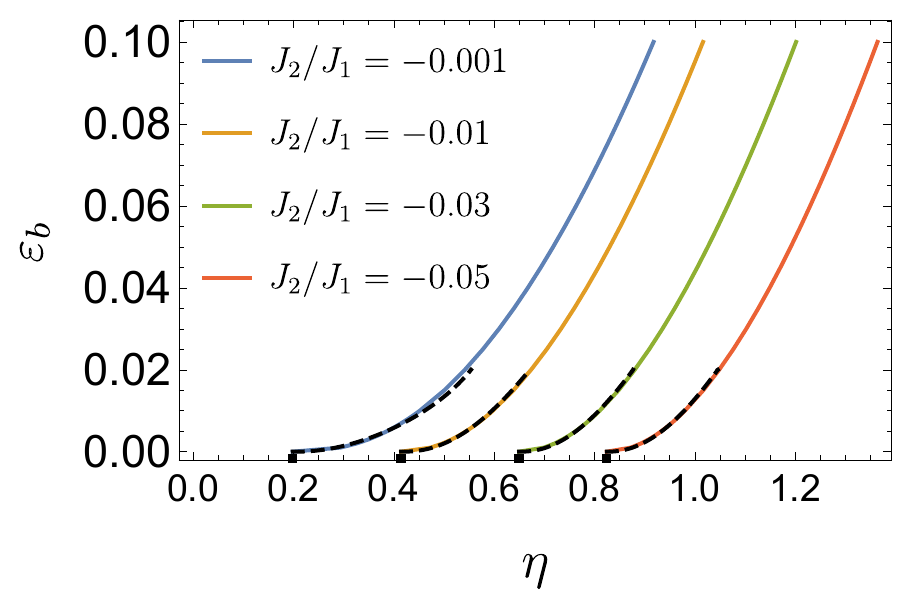}
	\caption{
Binding energy of the Potts-nematic excitons on the face-centered cubic (FCC) lattice as a function of interaction strength parameter $\eta  = J_i^z / J_i$.
Note that the bound state exists only at $\eta > \eta_c$ due to the peculiarity of a three-dimensional problem. Dashed lines demonstrate the asymptotics obtained by fitting the data with Eq.~\eqref{3D_solution_with_log}.}
	\label{fig:fcc_lattice}
\end{figure}

In the final part of this section we turn to an application of our theory to a model in three dimensions: the face-centric cubic (FCC) lattice. Our goal is to demonstrate that the analysis developed in Sec.~\ref{sec:theory} is not limited to two dimensions, but extends to three dimensions as well. The three-dimensional case is different in one important way, however. Since it is well known that bound states do not exist for arbitrarily shallow potential in three dimensions, making it a threshold phenomenon, we expect the two-magnon bound states to occur only for finite interaction between the magnons. In particular, this means that Eq.~\eqref{binding_energy} does not apply to the three-dimensional case. Below we describe how the analysis of Sec.~\ref{ssec:wc} can be adapted to the case of the FCC lattice.

We consider a $J_1$-$J_2$ model on the FCC lattice with antiferromagnetic $J_1$ and small ferromagnetic $J_2$. For a simple nearest neighbor FCC antiferromagnet ($J_2=0$), the Fourier transformed exchange coupling $J_{\bq}$ has minima on lines in momentum space. The lines of minima reflects the frustration of the FFC lattice and are defined by the constraint $(q_x,q_y)=(2\pi,0)$ with arbitrary $q_z$ (all symmetry equivalent constraints also define minima). Introducing ferromagnetic coupling between the next-nearest neighbors ($J_2 < 0$) lifts that degeneracy and results in minima of $J_{\bq}$ at the three wave vectors $\bQ_1= (2\pi,0,0)$, $\bQ_2=(0,2\pi,0)$, and $\bQ_3=(0,0,2\pi)$. These wave vectors satisfy Eq.~\eqref{eq:Q-relation}. 
The fact that there are three equivalent minima points implies that $\mathds{Z}_3$ symmetry is broken in the ordered phase with the ordering wavevector provided by one of these minima.
Such ordering is formed by stripes, which now can take one of the three directions parallel to the principal axes in the three-dimensional space.
We thus expect that in this case also one might expect intermediate Potts-nematic state at a range of single-ion anisotropy parameter $D$ between the ordered and paramagnetic phases.

We start with the demonstration of the solution for the binding energy $\varepsilon_b$, obtained by numerical integration in \eqref{M_definition}.
It is shown in Fig.~\ref{fig:fcc_lattice}, where $\varepsilon_b$ is plotted against the interaction strength $\eta = \eta_i = J_i^z / J^i$ for a number of $J_2 / J_1$ values.
As predicted above, we observe that the bound state is formed only at $\eta > \eta_c$, which is the main distinction with respect to the two-dimensional case.
At $\eta \gtrsim \eta_c$, the binding energy has approximately quadratic dependence on $\eta - \eta_c$. A more accurate formula \eqref{3D_solution_with_log} contains a logarithmic correction and is discussed below.
We plot it in dashed in Fig.~\ref{fig:fcc_lattice} and observe that it captures well the numerical dependence at small $\eta - \eta_c$ (with fitted parameters).

\begin{figure}
	\includegraphics[width=0.95\linewidth]{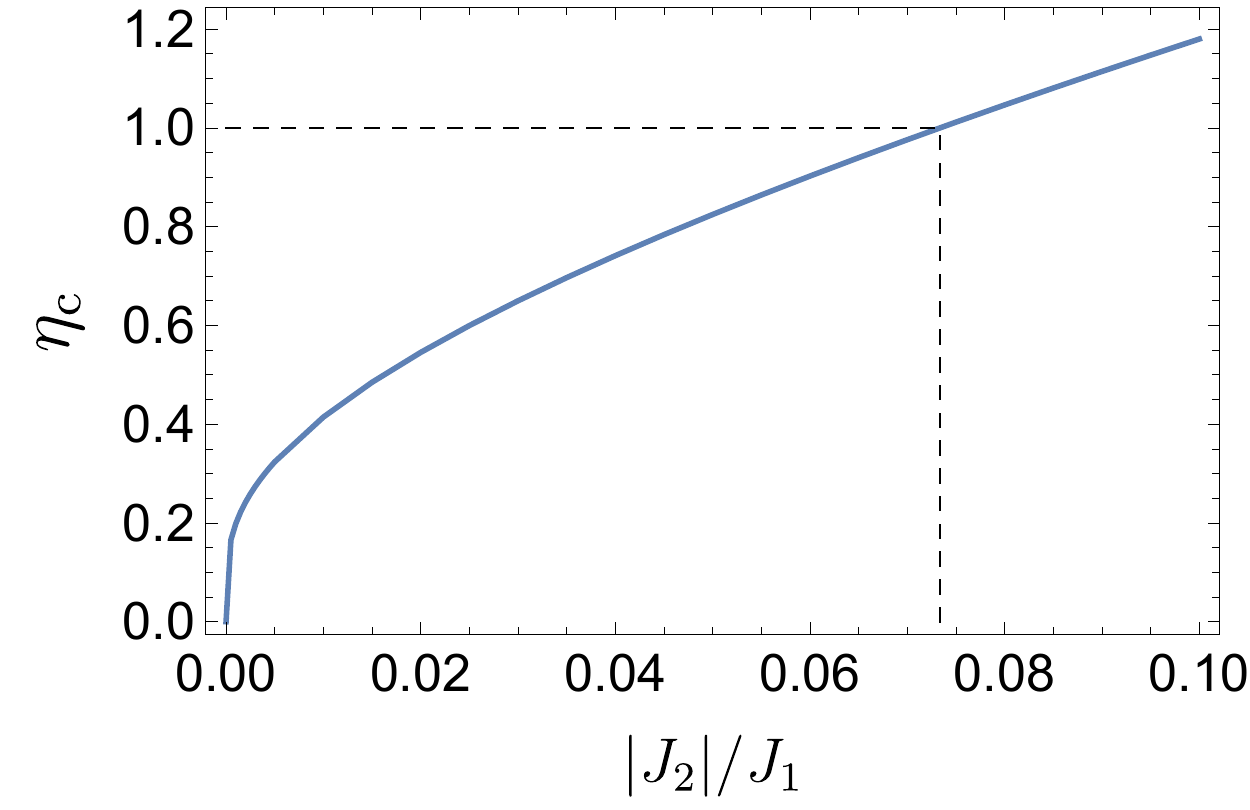}
	\caption{Critical values of the interaction parameter $\eta_c$ on the face-centered cubic (FCC) lattice as a function of spin couplings ratio $|J_2| / J_1$. 
Formation of Potts-nematic excitons and, hence, the vestigial phase occurs only at $\eta  = J_i^z / J_i > \eta_c$ due to the corresponding property of the three-dimensional potential well problem. 
}
	\label{fig:critical_zeta}
\end{figure}

Having identified the threshold nature of the bound state formation, we studied the dependence of the critical interaction $\eta_c$ on the spin-spin couplings (Fig.~\ref{fig:critical_zeta}).
The resulting dependence is monotonous and drops to zero in a sharp nonanalytical way at $J_2 \rightarrow 0$. 
This behavior is explained by the fact that at $J_2 = 0$ the single-magnon dispersion reaches its minimum on a set of lines, which effectively renders the problem two-dimensional.
From a quantitative perspective, it is interesting that for isotropic Heisenberg model with $\eta = J_i^z / J_i = 1$, the bound states exist only at $|J_2| / J_1 \lesssim 0.1$.

We proceed with the weak coupling theory for the three-dimensional case. Naive attempt at taking the integral in \eqref{M_definition} using expansion \eqref{final_eq_expand} in the vicinity of $\bQ_i$ points leads to a linearly divergent integral due to increased phase volume in three dimensions. The common resolution would be to subtract from the integrand of \eqref{final_eq_expand} its value at  $\varepsilon_b = 0$ (to be taken numerically over the whole Brillouin zone) and then proceed with the regularized integral. In the regime of small binding energies $\varepsilon_b$ that leads to the following implicit expression for $\varepsilon_b (\eta)$:
\be
	\eta^{-1} = \eta_c^{-1} -  B \, \sqrt{\frac{\varepsilon_b}{J_1}}.  \label{3D_solution}  
\ee
where $B$ is a constant determined by the density of states in the vicinity of $M$-points. 
The result indeed signifies that the bound state exists only at $\eta > \eta_c$. 
However, we found that Eq.~\eqref{3D_solution} is suitable to describe the asymptotics of the exact solution only at very small values of $\varepsilon_b$, which correspond to a much smaller scale than the one in Fig.~\ref{fig:fcc_lattice}. 
This observation is explained by the following considerations.
As discussed in the beginning of this subsection, at $|J_2| \rightarrow 0$ the dispersion acquires a very soft direction near the minimum, which eventually transforms into a line of degenerate minima.
This process leads to the increase of the single-magnon density of states, which, as discussed in previous sections, enhances the effect.
Given the smallness of relevant $J_2$ (in Fig.~\ref{fig:fcc_lattice} $|J_2| / J_1 \lesssim 0.1$) that means that at moderate $\varepsilon_b$ the value of \eqref{M_block} is not captured well by the quadratic expansion \eqref{final_eq_expand}. 
Instead, one can expect logarithmical corrections to \eqref{3D_solution} due to the nascent crossover to 2D.
Indeed, we find that a corrected formula
\be
	\eta^{-1} = \eta_c^{-1} -  B \, \sqrt{\frac{\varepsilon_b}{J_1}} \log \frac{\Lambda}{\varepsilon_b}  \label{3D_solution_with_log} 
\ee
captures the numerical dependence much better (see dashed lines in Fig.~\ref{fig:fcc_lattice}).

To conclude, our consideration predicts that the effect of exciton formation and appearance of the vestigial phase is weaker in three dimensions for nonspecific values of the spin-spin interactions.
However, in the vicinity of special points in the phase diagram ($J_2 = 0$ in our case), the effect becomes more pronounced due to the enhanced density of states of relevant magnons.


\section{Connection to experiment \label{sec:expts}}

The analysis presented in the previous section indicates that the instability towards a Potts-nematic phase is rather general and does not depend on the specific realization of the proposed model on any specific lattice. This generality motivates the important question whether material realizations of Potts-nematic quantum magnets exist among experimentally known material systems. In pursuit of this question, in this section we highlight and discuss a number of promising candidate materials to which our theory applies or relates. 

For our theory to directly apply, candidate materials should satisfy three requirements. First, the interactions between the spin must be such that magnetic order with broken $\mathds{Z}_3$ crystal symmetry is favored. This implies a tacit  constraint on the symmetry of the crystal structure and limits the search for materials to hexagonal, trigonal, tetrahedral, or octahedral systems. In practice, the search should be focused on materials or families of materials in which rotation symmetry broken magnetic order is observed, which is clearest signature of the nature of the exchange couplings. Examples will be discussed below. The second requirement is that the candidate materials exhibit significant easy-plane single-ion anisotropy, which implies that spin-orbit coupling is important. In our model, the single-ion anisotropy controls the (quantum phase) transition between the paramagnet and the ordered state, and therefore would be responsible for driving a magnetically ordered state into a putative Potts-nematic phase. Closely related to this second requirement is the third requirement: the quantum spins must be $S=1$ degrees of freedom, such that in the limit of large single-ion anisotropy the ground state is a perfect quantum paramagnet product state. In principle other integer-spin systems may be considered, such as $S=2$, since these also admit a  quantum paramagnetic ground state. 

Given these requirements, one of the more promising material classes is the family of transition-metal phosphorous trichalcogenides ($M$P$X_3$) 
\cite{Wildes:1998p6417, Wildes:2012p416004, Wildes:2015p224408, Lancon:2016p214407, Calder:2020p024408, Kim:2020p184429, Calder:2021p024414, Ni:2022p3283}. These are layered van der Waals materials in which the transition metal $M=$(Fe, Ni, Mn, Co) sites are magnetic and form a honeycomb lattice within each layer. The $X$ are occupied by chalcogen atoms, typically sulfur (S) or selenium (Se). The $M$P$X_3$ materials have attracted much attention recently as a versatile platform for exploring intrinsic two-dimensional magnetism in few-layer or monolayer systems. While some members of this class show antiferromagnetic Ne\'el order, others were found to realize antiferromagnetic zigzag order---precisely the type of order discussed in Sec.~\ref{ssec:honeycomb}. Furthermore, strong easy-axis or easy-plane behavior was observed, such as Ising-like easy-axis behavior in FePS$_3$ \cite{Wildes:2012p416004,Lancon:2016p214407,Coak:2021p011024} and XY-like easy-plane behavior in NiPS$_3$~\cite{Wildes:2015p224408,Lancon:2018p134414}. To explain the observed magnetic orderings across the family of $M$P$X_3$ compounds a $J_1$--$J_2$--$J_3$ honeycomb lattice spin model with single-ion anisotropy was proposed \cite{Wildes:2012p416004,Sivadas:2015p235425,Wildes:2015p224408,Chittari:2016p184428,Gu:2019p165405}, which, quite remarkably, is identical to the honeycomb lattice model studied in this work. It is for this reason that the  trichalcogenide magnets offer a particularly compelling venue for Potts-nematicity associated with rotation symmetry broken magnetic order.

Within the family of $M$P$X_3$ materials NiPS$_3$ deserves special attention. Not only do the ordered moments of the observed zigzag phase lie in the plane~\cite{Wildes:2015p224408,Lancon:2018p134414}, suggesting easy-plane anisotropy with $D>0$, but the Ni$^{2+}$ ions also give rise to $S=1$ spins on the honeycomb sites. This implies that NiPS$_3$ provides a realization of the model studied in this work.

It is important to point out one caveat concerning the bulk $M$P$X_3$ materials, in particular concerning the stacking of the constituent layers in the bulk structures. In the sulfur-based compounds $M$PS$_3$ with $M=$(Fe,Ni,Mn) the layers are stacked in monoclinic fashion, giving rise to space group $C2/m$. This means that the threefold rotation symmetry of the constituent honeycomb layers is already broken in the bulk structure, thus precluding the spontaneous breaking of rotation symmetry by the magnetic interactions. The coupling between the layers is believed to be very weak, however, and may be weak enough to prevent a selection of the nematic director by the monoclinic stacking direction. Indeed, encouraging evidence for this has recently been reported by linear and non-linear optical measurements \cite{Ni:2022p3283}, which have found three different magnetic domains below the Ne\'el temperature within one homogeneous structural domain.

A second material of interest is the iron-intercalated transition metal dichalcogenide (TMD) $\text{Fe}_x \text{Nb} \text{S}_2$ with $x = 1/3$, which is a member of the class of intercalated TMDs $M_x T A_2$~\cite{Friend:1977p1269,Parkin:1980p65,Mankovsky:2016p184430}. Here $M$ is a transition metal, $T$=(Ta, Nb), and $A$=(S, Se). The magnetic Fe atoms are intercalated between the adjacent van der Waals layers of $2H$-NbS$_2$ and form a triangular lattice with $\sqrt{3}\times\sqrt{3}$ periodicity. The two neighboring triangular lattice layers of Fe atoms that make up the unit cell (and are located at $c=1/4$ and $c=3/4$) are shifted with respect to each other, effectively forming a honeycomb structure. The resulting space group of the intercalated structure is $P6_322$. Importantly, both optical birefringence \cite{Little:2020p1062,Wu:2022p021003} and detailed neutron crystal diffraction \cite{Wu:2022p021003} measurements have reported observation of rotational symmetry broken stripe order. Here, the two triangular lattice layers each form the ordering pattern shown in Fig.~\ref{fig:triangular} and together form the honeycomb lattice stripy order shown in Fig.~\ref{fig:honeycomb_results}. As such, the observed magnetic order falls in the class of orderings considered in our work. (Note that Ref.~\onlinecite{Wu:2022p021003} reported great sensitivity of the ordering pattern to small changes in the intercalation parameter $x$ close to the commensurate value $1/3$.) Furthermore, the Fe$^{2+}$ ions give rise to $S=2$ spins  \cite{Sundararajan:1983p773,Gorochov:1981p621}. To understand the observed magnetic orderings, recent theory works proposed a Heisenberg spin with extended exchange interactions and single-ion anisotropy~\cite{Haley:2020p043020,Weber:2021p214439}, essentially equivalent to our honeycomb lattice model. An open question is the nature of the single-ion anisotropy, however, with initial estimates suggesting it might be negative, thus implying an easy-axis anisotropy~\cite{Haley:2020p043020,Weber:2021p214439}.

We conclude this section with a general remark. To obtain the nematic phase at low temperatures one needs a suitable tuning knob to weaken and eventually destroy the magnetic order. This can be provided by varying strain and pressure in the system.  An alternative possibility is the formation of Potts-nematic phase at the thermal phase transition as function of temperature, rather than the quantum phase transition studied here. A theory of thermal transition to the Potts-nematic phase will be reported elsewhere.

\section{Discussion and Conclusion}
\label{sec:discussion}

The central result of this paper is the demonstration that a magnetically disordered Potts-nematic phase can exist as an intermediate phase at the quantum phase transition from a quantum paramagnet to a $\mathds{Z}_3$ threefold rotation symmetry broken antiferromagnet. Since such a phase is magnetically disordered, spin-rotation symmetry and time-reversal symmetry are preserved, yet a $\mathds{Z}_3$ crystal rotation symmetry is broken, thus giving rising to nematic order. The relevant order parameter for the nematic phase is expressed in terms of bilinears of the primary magnetic degrees of freedom, i.e., the spin variables.

Our conclusion is inferred from the analysis of two-magnon excitations in the paramagnetic state of an $S=1$ quantum spin model, in which the transition between the paramagnet and the magnetic state is regulated by the value of easy-plane single-ion anisotropy. With both numerical and analytical methods we established the existence of two-magnon bound states---``excitons''---with the angular momentum $l=2$. That makes the two-particle gap close slightly before the single-particle gap as the transition is approached, leading to the proliferation of excitons. The latter can be described by a Jastrow ansatz \cite{Zaletel:2014p155142,Wang:2017p184409}.  We derived a general equation for the excitonic bound state in terms of lattice harmonics [Eqs.~\eqref{det_zero} and \eqref{M_block}] and derived its analytical solution for the weak-coupling regime \eqref{binding_energy}.
The latter expresses the binding energy and, hence, the width of the vestigial phase, as a function of magnon density of states at the dispersion minima (in the paramagnetic state) and the effective interaction strength, which is a linear combination of the spin-spin coupling constants. Analytical perturbation theory is corroborated by a detailed numerical analysis for the cases of triangular, honeycomb, and face-centered cubic lattices (see Figs.~\ref{fig:triangular_results}--\ref{fig:critical_zeta}).

It is important to describe the relation between the character of the vestigial phase determined by the two-magnon wavefunction and the properties of the single-magnon dispersion above the paramagnetic phase. 
In the present paper we assumed that the dispersion has three degenerate minima in the Brillouin zone ($Q_1$, $Q_2$, and $Q_3$ momenta, satisfying $2 Q_i \sim 0$), which are related to each other by an in-plane rotation for the cases of triangular and honeycomb lattice.
Under these assumptions, perturbative expression \eqref{lambda_general} for the effective coupling constants in different angular momenta channels ($l$) nullifies for $l=1$ and $l=3$ leaving only Potts-nematic channel with $l=2$ and unphysical symmetric wavefunction with $l=0$. 
Similarly, we checked that in the setup of Ref.~\onlinecite{Wang:2017p184409}, in which the dispersion had two inequivalent minima, the only physical channel arising in the perturbation theory is the chiral one ($l=3$). 
This relation provides a universal scheme allowing to determine the nature of the vestigial phase given the single-particle spectrum. 
In the numerical calculation we observed that the effective couplings in the `wrong' angular momentum channels are nonzero, but they are small and thus irrelevant deep inside the domain with a given magnetic ordering [see Fig.~\ref{fig:triangular_results}(b)].

While in the present work we focused on the specific case of an $S=1$ spin model, our analysis straightforwardly generalizes to any integer spin system. Indeed, the ground state at large values of the single-ion anisotropy in that case is still a perfect paramagnet, given by a product of $S^z = 0$ states on each site. A spin flip to a state with $S^z = s$ on any site costs an energy $D s^2$ and therefore, the energetically lowest excitations in this case still correspond to $s = \pm 1$. It means that for the purpose of determining the instability of the paramagnet marked by a gap closure, one can restrict the consideration to these two branches only, which makes all further analysis the same as in the $S=1$ scenario. In contrast, we expect that half-integer models behave qualitatively different, since they do not admit a paramagnetic product state.

The prediction of the formation of Potts-nematic phase at low temperatures in the vicinity of the quantum phase transition poses a natural question: does there exist a similar vestigial phase at the thermal phase transition between a striped antiferromagnet and the symmetry high-temperature phase? This question will be addressed in a separate paper.

We conclude by emphasizing that there exist a number of different materials which provide promising venues for experimental observation of Potts-nematicity in quantum magnets. As discussed in Sec.~\ref{sec:expts}, given its currently established magnetic properties, the $M$P$X_3$ family of transition-metal trichalcogenides is of particular interest. Within this class of materials NiPS$_3$ is most promising, as it is likely to provide a realization of the model considered in this work. Another possible candidate material is the $\text{Fe}$-intercalated dichalcogenide $\text{Nb} \text{S}_2$.
\vspace{0.3cm}

\section{Acknowledgements}

We gratefully acknowledge useful discussions with R. M. Fernandes, A. V. Chubukov, and L. Wu. This research was supported by the National Science Foundation Award No. DMR-2144352.

\bibliography{nematic_excitons}

\clearpage

\appendix

\section{Generalization to non-Bravais lattices \label{app:gen}}

Here we describe the generalization of the two-magnon problem to the case of a non-Bravais lattice, which requires a proper account for the sublattice degree of freedom. We start we the expression for the $XXZ$ Hamiltonian of the form
\begin{multline}
	\mathcal{H} = \frac{1}{2} \sum_{\br, \bdelta, \alpha, \beta} (J_{\bdelta})_{\alpha\beta}  (S_{\br\alpha}^{x}S_{\br + \bdelta,\beta}^{x} + S_{\br\alpha}^{y}S_{\br + \bdelta,\beta}^{y})  \\
+  \frac{1}{2} \sum_{\br, \bdelta, \alpha, \beta} (J^z_{\bdelta})_{\alpha\beta} S_{\br\alpha}^{z}S_{\br + \bdelta,\beta}^{z},
\end{multline}
where $\br$ and $\br+\bdelta$ are Bravais lattice vectors and $\alpha,\beta$ label the spins within the unit cell, i.e., the sublattice degrees of freedom. 
Two-magnon wavefunction $\psi_{\br, \alpha; \br' \beta}$ also carries sublattice indices. 
For the clarity of the narration, we will consider zero center-of-mass momentum excitons (generalization to the opposite case is straightforward).
Then, the Fourier transform of the wavefunction reads
\be
	\psi_{\br, \alpha; \br' \beta} = \sum_{\bq} \psi_{\alpha \beta} (\bq) e^{i\bq \cdot (\br + \bfell_{\alpha} - \br' - \bfell_{\beta})},
\ee
where $\bfell_{\alpha}$ determines the position of $\alpha$-sublattice site inside a unit cell. 

We proceed with projecting the Schroedinger equation to the subspace of two opposite-spin magnons as was done in the main part [Eq.~\eqref{integral_equation}].
It is convenient to interpret $\psi_{\alpha \beta} (\bq)$ as a matrix and write the variational Schroedinger equation in the matrix form:
\be
\label{Schroedinger_multiband}
	(E - 2D) \psi(\bq) - \hat{J}_{\bq} \cdot \psi(\bq) - \psi(\bq) \cdot \hat{J}_{\bq}^\dagger = \text{Int}[\psi].
\ee
Here $\hat{J}_{\bq}$ is the Fourier transform of the interaction matrix
\be \label{J_q_matrix_definition}
	(J_{\bq})_{\alpha\beta} = \sum_{\bdelta} (J_{\bdelta})_{\alpha\beta} e^{i\bq\cdot(\bdelta +\bfell_\beta-\bfell_\alpha)},
\ee
which reduces to the single-magnon energy $\varepsilon(\bq)$ in the Bravais-lattice case [see paragraph under \eqref{B_def}].
$\text{Int}[\psi]$ is the interaction term appearing due to magnon-magnon interaction, provided by the $S^z S^z$ part of the spin coupling. It has the following matrix elements:
\be
(\text{Int}[\psi])_{\alpha \beta} = - \sum_{\bdelta} (J_{\bdelta}^z)_{\alpha\beta} \ B_{\alpha, \beta, \bdelta}  \ e^{i\bq\cdot(\bdelta +\bfell_\beta-\bfell_\alpha)},
\ee
where $B_{\alpha, \beta, \bdelta} = \psi_{\br, \alpha; \br + \bdelta, \beta}$ is the real-space wavefunction amplitude at two magnons residing on $\alpha$ and $\beta$ sublattices in the unit cells separated by the Bravais lattice vector $\bdelta$. That is a correct definition because $\psi_{\br, \alpha; \br + \bdelta, \beta}$ does not depend on $\br$ in the case of zero center-of-mass momentum. Importantly, \eqref{J_q_matrix_definition} contains only the elements $B_{\alpha, \beta, \bdelta}$ which are multiplied by nonzero spin-spin couplings $(J_{\bdelta}^z)_{\alpha\beta}$. Thus, for a model with finite-range interaction, there is a finite number of $B_{\alpha, \beta, \bdelta}$ components. For instance, on a $J_1-J_2-J_3$ model on a honeycomb lattice  for each of the two choices for $\alpha$, there are three nearest neighbors, six next-nearest, and three third-nearest ones, constituting 12 elements for one $\alpha$ and 24 total components of $B_{\alpha, \beta, \bdelta}$.
To proceed, we need to express the latter via the momentum-space wavefunction:
\be
\label{B_definition}
	B_{\alpha, \beta, \bdelta} = \sum_{\bm{p}} e^{-i \tilde{\bdelta} \bm{p}} \, \psi_{\alpha \beta} (\bm{p}),
\ee
where we introduced the following notation for brevity: 
\be
	\tilde{\bdelta} = \bdelta +\bfell_\beta-\bfell_\alpha .
\ee

%

Next step is to diagonalize the l.h.s. of  \eqref{Schroedinger_multiband} by expressing $\psi_{\alpha \beta} (\bq)$ via the eigenstates $e_{a}(\bq)$ of $\hat{J}_{\bq}$ matrix:
\be
\label{psi_through_eigenvectors}
	\psi(\bq) = \sum_{ab} c_{ab} (\bq) e_{a}(\bq) e_{b}^{T}(-\bq), \quad \hat{J}_{\bq} e_{a}(\bq) = \varepsilon_{\bq}^{(a)}  e_{a}(\bq),
\ee
where $a$ enumerates the Bloch bands of magnons. 
Then, \eqref{Schroedinger_multiband} takes the form 
\be
\label{Scroedinger_multiband_diagonalized}
	\sum_{ab} \left[E - 2D - \varepsilon_{\bq}^{(a)} - \varepsilon_{-\bq}^{(b)} \right] c_{ab} \, e_{a}(\bq) e_{b}^{T}(-\bq) = \text{Int}[\psi],
\ee
Assuming that $e_{a}(\bq)$ constitute an orthonormal basis for each $\bq$ and multiplying \eqref{Scroedinger_multiband_diagonalized} by $e_{a'}^{\dagger}(\bq)$ from the left and by $e_{b'}(\bq)$ from the right, we get $c_{ab}$:
\be
\label{c_coefs}
	c_{ab}(\bq) = - \sum_{\alpha, \beta, \bdelta} \frac{ (J_{\bdelta}^z)_{\alpha\beta} \, e^{i \tilde{\bdelta} \bq} \cdot e_{a}^{\dagger}(\bq)  \, B_{\alpha, \beta, \bdelta} \, e_{b}^{*}(-\bq) }{E - 2D - \varepsilon_{\bq}^{(a)} - \varepsilon_{\bm{-q}}^{(b)} } .
\ee
We then substitute \eqref{c_coefs} into \eqref{psi_through_eigenvectors} and further into \eqref{B_definition} to find the self-consistency equation
\be \label{equation_for_B_nonBravais}
	B_{\alpha, \beta, \bdelta} = \sum_{\alpha', \beta ', \bdelta '} \mathcal{M}_{\alpha, \beta, \bdelta; \alpha', \beta', \bdelta '} B_{\alpha', \beta', \bdelta '},
\ee
where
\be
\label{M_multiband}
	\mathcal{M}_{\alpha, \beta, \bdelta; \alpha', \beta', \bdelta '} = \sum_{{\bq},a,b}  \frac{ (J_{\bdelta '}^z)_{\alpha' \beta'} e^{i (\tilde{\bdelta}' - \tilde{\bdelta})\bq} \ P_{\alpha \alpha'}^{(a)} (\bq) P_{\beta \beta'}^{(b)}(-\bq)}{2D + \varepsilon_{\bq}^{(a)} + \varepsilon_{\bm{-q}}^{(b)} },
\ee
where $P^{(a)} (\bq) = e_{\bq}^{(a)} [e_{\bq}^{(a)}]^{\dagger}$ is the projector to the $a$th band acting in the sublattice space, and $E=0$ value was taken since we are looking for the instability, which arises when the excitons become gapless.

Condition for \eqref{equation_for_B_nonBravais} to have solution is given by the same equation $\det \mathcal{M} - \mathbb{1} = 0$ as in the main text [Eq.~\eqref{det_zero}].
However, expression \eqref{M_multiband} for $\mathcal M$ is slightly more involved and includes projectors to magnon bands and summation over the latter.
Nevertheless, for an isotropic model one can still block diagonalize $\mathcal M$ with the help of lattice harmonics and proceed with the numerical computation.
In Section \ref{ssec:honeycomb} we demonstrate how that works for the case of the honeycomb lattice. 

Weak coupling theory for a non-Bravais lattice is regulated by the parameters $(\eta_{\bdelta})_{\alpha \beta} = (J_{\bdelta}^z)_{\alpha \beta} / (J_{\bdelta})_{\alpha \beta}$, which determine interaction between the single-magnon excitations. 
In an isotropic model the couplings depend only on the relative distance $\tilde{\bdelta}$ between the sites:
\be
	(J_{\bdelta})_{\alpha \beta} = J(|\bdelta +\bfell_\beta-\bfell_\alpha|).
\ee
For simplicity, we would also assume that $\eta$ is the same for all couplings so that the theory has just one perturbation parameter. 
At small $\eta$, which corresponds to a shallow well problem, common knowledge is to expect the exciton binding energy to be small, while its size to be large in real space and small in reciprocal space.
That allows to consider only the low-lying excitations in the summation in \eqref{M_multiband}, which boils down to considering only the lower band and expanding the dispersions $\varepsilon_{\bm{q}}^{(a)}$ in the vicinity of their minima, which we denote as $\bQ_i$. Moreover, we will evaluate the numerator of \eqref{M_multiband} strictly at $\bQ_i$ points. Thus, we rewrite the summation in the vicinity of each minimum in terms of $\bm{p} = \bq - \bm{Q}_i$. 
\be
\label{M_multiband_lowest}
	\mathcal{M}_{\alpha, \beta, \bdelta; \alpha', \beta', \bdelta '} = \sum_{i} \int\frac{d \bp}{V_{BZ}}  \frac{ (J_{\bdelta '}^z)_{\alpha' \beta'} e^{i (\tilde{\bdelta}' - \tilde{\bdelta}) \bQ_i} \ P_{\alpha \alpha'}^{\bQ_i} P_{\beta \beta'}^{\bQ_i}}{\varepsilon_b + \bp^T A_i \bp },
\ee
where $P_{\alpha \beta}^{\bQ_i} = P_{\alpha \beta}^{(\text{low})} (\bQ_i)$ are projectors onto the lowest magnon band, binding energy $\varepsilon_b = 2D + \varepsilon_{\bQ_i} + \varepsilon_{-\bQ_i}$, and 
\be
	(A_i)_{mn} = \left.  \frac{ \partial^2 (\varepsilon_{\bq}^{(\text{low})} + \varepsilon_{\bm{-q}}^{(\text{low})}) } { \partial q_m\partial q_n}  \right|_{\bQ_i} 
\ee
is the expansion of the magnon dispersion in the vicinity of the global minima $\bQ_i$. 
Eq.~\eqref{M_multiband_lowest} has the same general form as the analogous equation for a Bravais lattice \eqref{final_eq_expand}. 
For a quadratic magnon band minima one can proceed with integrating \eqref{M_multiband_lowest} over $\xi = \bp^T A_i \bp$ as was done in \eqref{final_eq_log}. 
The corresponding density of states $\nu_{\bQ_i}$ should contain only the contribution of the lowest band at $\bQ_i$ point. 
Such integration leads to the same functional dependence as in \eqref{final_eq_log} with some value of $\lambda$ and density of states to be understood as explained above.
For the case of the honeycomb lattice, which has dispersion minima at M-points, these quantities are given by \eqref{lambda_honeycomb} and \eqref{dos_honeycomb}.
We discuss the result of the weak-coupling approach for this case  in detail in Section~\ref{ssec:honeycomb}.

\end{document}